\documentclass[pra,twocolumn,superscriptaddress,nofootinbib]{revtex4-1}
\usepackage{graphicx}
\usepackage{bm}
\usepackage{amssymb,amsmath,mathrsfs}
\usepackage{color}
\usepackage{amstext}
\usepackage[english]{babel}
\usepackage{latexsym}
\usepackage{array}   
\usepackage{multirow}         
\usepackage[usenames,dvipsnames]{xcolor}
\usepackage[colorlinks=true,citecolor=Blue,linkcolor=RubineRed,urlcolor=Blue]{hyperref}
\usepackage[all]{xy}     
\usepackage[usenames,dvipsnames]{xcolor}

\definecolor{mygreen}{RGB}{0,110,62}
\definecolor{mypurple}{RGB}{134,3,191}

\begin{document}

\title{Exact dynamics and qubit inversion of  non-Hermitian driven two-level systems}
\author{Ivan~A. Bocanegra-Garay\href{https://orcid.org/0000-0002-5401-7778}{\includegraphics[scale=0.45]{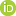}}}
\email[]{ivanalejandro.bocanegra@uva.es}
\affiliation{Departamento de F\'isica Te\'orica, At\'omica y \'Optica and  Laboratory for Disruptive Interdisciplinary Science, Universidad de Valladolid, 47011 Valladolid, Spain}
\author{Luis M. Nieto\href{https://orcid.org/0000-0002-2849-2647}{\includegraphics[scale=0.45]{orcid}}}
\email[]{luismiguel.nieto.calzada@uva.es}
\affiliation{Departamento de F\'isica Te\'orica, At\'omica y \'Optica and  Laboratory for Disruptive Interdisciplinary Science, Universidad de Valladolid, 47011 Valladolid, Spain}

\date{\today}

\begin{abstract}
The supersymmetric structure of a \textit{generalized} non-Hermitian driven two-level system is demonstrated. A unitary rotation turns the Hamiltonian into a more convenient form. After decoupling a set of differential equations, the supersymmetric structure of the problem can be unequivocally appreciated. Performing a spectral analysis of an auxiliary stationary Schr\"odinger-like equation, complex time-dependent driving functions are obtained for which the corresponding (time-dependent) Schr\"odinger equation can be straightforwardly solved. Such complex drivings are seen to produce transitions in the qubit state in different, however interesting, manners. We believe that the results reported here will be of interest for designing and carrying out various experiments in laboratories specializing in nuclear magnetic resonance or in optics with gain and loss materials.
\end{abstract}
\maketitle

\section{Introduction}
The semi-classical Rabi model for a qubit \cite{Rabi1936,Rabi1937}, i.e. the Schr\"odinger equation for the Hamiltonian
\begin{equation}\label{Rabi}
    \mathcal{\hat H} = \Delta\hat\sigma_z + w(t)\hat\sigma_x,
\end{equation}
has been the subject of active interest in research in the last decades \cite{Braak2016}. The real constant $\Delta$ in (\ref{Rabi}) represents the difference between the two energy levels of the qubit (or two-level system) and  $w(t)$ represents a time-dependent driving function stimulating transitions between the two qubit energy levels. The case of a classical harmonic electric field $w(t) = \cos(\omega t)$ driving the two-level system was analyzed \cite{Moiseyev2009,Moiseyev2011,Lee2015,Castanos2019}. However, exact solutions associated with time-dependent Hamiltonians of the type  (\ref{Rabi}) remain rare (for some exceptions, see \cite{Gangopadhyay2010,deClercq2016,Giscard2020}).

Non-Hermitian versions of the system (\ref{Rabi}) have also been considered in the literature \cite{Bender2007,Moiseyev2011,Ibanez2011,Torosov2013,Zheng2013,Lian2014,DAmbroise2014,Gong2015,Lee2015,Wu2017,Hassan2017,Xie2018} (see also \cite{Pires2021}). Non-Hermitian (driven) qubits have attracted attention in recent years \cite{Dogra2021,Xu2022,Leng2023,Kivela2024,Pan2024,Jabraeilli2025,Liu2025,Fan2025}. The non-Hermitian $\mathcal{PT}$-symmetric Rabi model obtained if $w(t) = i\cos(\omega t)$ in \eqref{Rabi} has been studied in detail \cite{Moiseyev2011,Liu2025}, and is said to be $\mathcal{PT}$-symmetric because this Hamiltonian 
commutes with the product of the parity operator $\mathcal{P} = \hat\sigma_z$ and the time-reversal operation $\mathcal{T}$ (complex conjugation): $[\mathcal{PT},\mathcal{\hat H}] = 0$ \cite{Bender1998} (see also \cite{Ruter2010,Liu2024} for $\mathcal{PT}$-symmetric systems in a different context). 

Supersymmetric quantum mechanics (SUSY QM) is a valuable technique for finding new exactly solvable quantum potentials, for instance, in the context of one-dimensional stationary Schr\"odinger equations, starting from given exactly solvable quantum potentials \cite{Cooper1995,fernandez1999,Mielnik2000,Samsonov2004,Correa2007,miri2013,Correa2015,CruzyCruz2020,Garcia2023} (free-particle, harmonic oscillator or Lam\'e). See also \cite{Zuniga2014,Schulze2017,Contreras2019,Maldonado2021,Kafuri2024,BocanegraGaray2024,Bocanegra_2024PRR} for SUSY transformations in miscellaneous contexts. In particular, exactly solvable one-dimensional complex potentials can be constructed from exactly solvable real ones \cite{RosasOrtiz2015,RosasOrtiz2018,Zelaya2020,Bocanegra2022}.

In this work, we use a SUSY QM approach \cite{Bender1998SUSY} to construct complex time-dependent coupling functions $w(t)$ for a non-Hermitian time-dependent semi-classical Rabi model of the type  (\ref{Rabi}). We do this by bringing the Hamiltonian into a suitable form via a unitary transformation, and identify the resulting system of two differential equations (after decoupling) as a SUSY pair of stationary Schr\"odinger-like equations \cite{Cooper1995,fernandez1999,Mielnik2000,Samsonov2004,Correa2007,Correa2015,CruzyCruz2020}. By fixing the potential of one of the stationary Schr\"odinger equations, the coupling function $w(t)$ is fixed as well, and the solutions of the time-dependent Schr\"odinger equation for the Hamiltonian (\ref{Rabi}) are straightforwardly obtained.
In this paper we elucidate exciting features of the system under consideration. For instance, we show that when the coupling function $w(t)$ is purely imaginary, the transitions between the two energy levels are suppressed. In contrast, when $w(t) = w_R(t) + iw_I(t)$, with $w_R\neq 0$ and $w_I$ real functions of time, the qubit might perform transitions. Both periodic and nonperiodic transitions appear when $w_{R,I}(t)\neq 0$, for some interval of $t$. Thus, transitions are possible even in the presence of a nontrivial imaginary part of the coupling function, i.e. $w_I(t)\neq 0$.

Non-Hermitian physics has gained exceptional importance, since real physical processes are, in fact, nonconservative. Nevertheless, as mentioned, there are only a few time-dependent non-Hermitian exactly-solvable systems in the literature. In this framework, the present paper aims to fill that gap in an elegant way by using the simplest approach: we take advantage of the form of the differential equations that govern the evolution of the non-Hermitian time-dependent system and the available knowledge on one-dimensional SUSY QM in stationary systems.

This work is organized as follows.  
In Section~\ref{sec.model}, the model under examination is presented: the SUSY structure of the decoupled system of differential equations defining a unitary rotation of the model under investigation is reviewed; in particular, the construction of complex \textit{superpotentials} (coupling functions) is considered. 
In Section \ref{sec.freepotential}, the simplest example of the general SUSY approach is examined. 
In Section \ref{sec.numerical}, some numerical experiments are presented to show the applicability of the results obtained in the previous sections.
In Section \ref{sec.experiment}, the experimental feasibility of the proposed theoretical model is discussed. 
Finally, in Section \ref{sec.conclusions}, the main conclusions of the manuscript are presented, and some future work is proposed.

\section{Exactly-solvable model}\label{sec.model}

\begin{figure}[t!]
\centering
\includegraphics[width=1\linewidth]{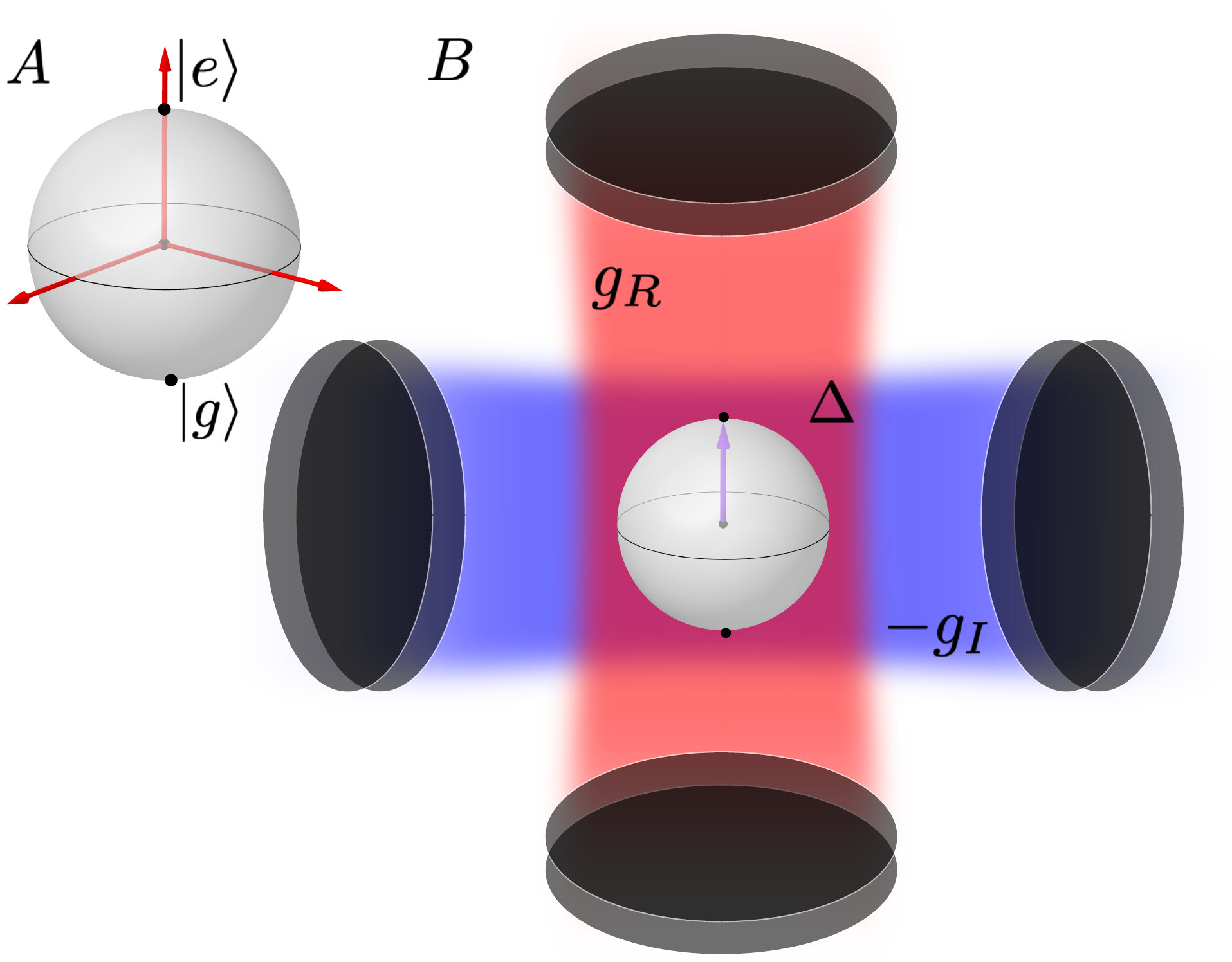}
\caption{\label{fig_gainloss}  {\bf Schematic of the non-Hermitian system under consideration.} In $A$, a qubit (two-level system) is represented by the Bloch sphere. In $B$, the qubit, with energy split $\Delta$, is subjected to a mechanism that produces transitions in it (blue shadow) and another that represents gain/loss  of the non-Hermitian system (red shadow). } 
\end{figure}

Let us consider a two-level system (Figure \ref{fig_gainloss}A) with energy split $\Delta = \omega_e - \omega_g$, where $\omega_{e,g}$ are the energies of the excited $|e\rangle$ and ground $|g\rangle$ states, respectively, subjected to a mechanism that stimulates transitions in the qubit, and to another one that acts as an external gain/loss (Figure \ref{fig_gainloss}B). The  Schr\"odinger equation ($\hbar = 1$) reads 
\begin{equation}\label{piti}
    \mathcal{\hat H} |\phi\rangle = i\partial_\tau |\phi\rangle, \qquad \mathcal{\hat H} = \Delta \hat\sigma_z + i g(\tau)\hat\sigma_x,
\end{equation}
where $\hat\sigma_z, \hat\sigma_x$ are Pauli matrices. Note that $\mathcal{\hat H}$ is a time-dependent ($\tau$) non-Hermitian Hamiltonian. The complex term  $ig(\tau)$ represents a time-dependent coupling between the qubit levels and also accounts for the gain/loss of the system. When $g(\tau)$ is restricted to be a real even function of time, the Hamiltonian (\ref{piti}) belongs to the family of $\mathcal{PT}$-symmetric Hamiltonians \cite{Bender1998,Ruter2010}, with $\mathcal{P} = \hat\sigma_z$ and $\mathcal{T}$ the complex-conjugation. Up to now, $g(\tau)$ is a complex function to be determined, and no further assumptions on it have been made. By defining a dimensionless time $t= \Delta\tau$, the following Schr\"odinger equation is obtained
\begin{equation}\label{dimensionless}
 \hat H_0|\phi\rangle = i\partial_t|\phi\rangle, \ \ 
    \hat H_0 = \hat\sigma_z + if(t)\hat\sigma_x,\ \  
    f(t) = {g(\tau)}/{\Delta}.
\end{equation}
Letting $|\phi\rangle = \exp(-i\frac{\pi}{4}\hat\sigma_y)|\psi\rangle$, (\ref{dimensionless}) transforms into
\begin{equation}\label{pete}
    \hat H |\psi\rangle = i\partial_t |\psi\rangle, \qquad \hat H = if(t) \hat\sigma_z - \hat\sigma_x.
\end{equation}
In the following, a simple method for solving the Schrödinger equation (\ref{pete}), developed on SUSY foundations, will be presented.

\subsection{SUSY structure of the problem}

By writting $|\psi\rangle=\left(\begin{array}{cc}
     a_1(t)  \\
     a_2(t) 
\end{array} \right)$, with $a_{1,2}(t)$ complex functions of $t$, (\ref{pete}) becomes the set of coupled differential equations
\begin{eqnarray}\label{coupled1}
 &&   i\dot a_1(t) = if(t) a_1(t) - a_2(t),
\\
  &&   i\dot a_2(t) = - if(t) a_2(t) - a_1(t),
  \label{coupled2}
\end{eqnarray}
where the dots represent derivation with respect to $t$. By differentiating, the following pair of uncoupled second-order stationary Schr\"odinger-like (in the variable $t$, not $x$) differential equations is obtained
    \begin{equation}\label{decouple}
    \hat H_1 a_1(t) = E a_1(t),
\quad
    \hat H_2 a_2(t) = E a_2(t), 
\end{equation}
with $\hat H_{1,2} = -\partial_{tt} + V_{1,2}(t)$, $E=0$ and where we have defined \cite{RosasOrtiz2015,RosasOrtiz2018}
\begin{equation}\label{susyp}
       V_1(t) = f^2(t) + \dot f(t) - 1,
\quad
       V_2(t) = f^2(t) - \dot f(t) - 1.
   \end{equation}
The SUSY structure of (\ref{decouple}) can be seen by identifying (\ref{susyp}) as two SUSY quantum potentials $V_1(t)$ and $V_2(t)$ (see \cite{RosasOrtiz2015,RosasOrtiz2018} and references therein), where
\begin{equation}\label{part}
    V_2(t) = V_1(t) - 2\dot f(t),
\end{equation}
connected by the coupling (called \textit{superpotential} in the SUSY QM jargon) or driving function defined by $f(t)$. 
When defining the operators
\begin{equation}\label{op}
    \hat A = -\partial_t - f(t), \qquad \hat B = \partial_t - f(t),
\end{equation} 
equations (\ref{decouple}) can be cast, respectively, in the form
\begin{equation}\label{SUSY}
    (\hat A \hat B + \epsilon) a_1(t) = 0, \qquad (\hat B \hat A + \epsilon) a_2(t) = 0,
\end{equation}
where $\epsilon = -1$. Thus, $a_2(t) \propto \hat B a_1(t)$ and $a_1(t)\propto \hat A a_2(t)$. Indeed, it is found that
\begin{equation}\label{propto}
    a_2(t) = -i \hat B a_1(t), \qquad a_1(t) = i \hat A a_2(t),
\end{equation}
which is consistent with \eqref{coupled1} and \eqref{coupled2}.

It is possible to take advantage of the SUSY pair in (\ref{decouple}) by choosing $V_1(t)$ and $V_2(t)$ as any known pair of SUSY partner potentials, for which the solutions $a_1(t)$ and $a_2(t)$ are known \cite{Cooper1995,fernandez1999,Mielnik2000,Samsonov2004,Correa2007,miri2013,Correa2015,CruzyCruz2020,Garcia2023}. This fixes the superpotential (coupling or driving function) $f(t)$ that connects the partner Hamiltonians via (\ref{part}).

The constant $\epsilon$ in both expressions of (\ref{SUSY}) is known as the \textit{factorization energy}. In usual SUSY approaches, the factorization energy $\epsilon$ is a free parameter \cite{Mielnik2000,RosasOrtiz2015,RosasOrtiz2018}. However, in this case, it is already fixed, by the form of the coupled equations \eqref{coupled1} and \eqref{coupled2}, as $\epsilon = -1$.

It is also interesting that a SUSY pair of equations similar to (\ref{decouple}) is given in equations (16) and (17) of Reference \cite{Gong2015}. However, there are some notable differences worth mentioning. 
First, the method used in Ref. \cite{Gong2015} to obtain such a pair of SUSY equations is somewhat cumbersome; in our approach it is a natural consequence of the change of variable indicated above (\ref{pete}). 
Second, in \cite{Gong2015} the authors specifically consider a periodic modulation; in our approach an explicit form of $f(t)$ has not been assumed. In fact, we will consider both periodic and nonperiodic modulation functions $f(t)$.

\subsection{Complex driving functions (superpotentials)}

The superpotential $f(t)$ defining the coupling between the two qubit levels can be taken as a real function. If such a function is even, the Hamiltonian (\ref{piti}) is $\mathcal{PT}$-symmetric, with properly defined $\mathcal{P}$ and $\mathcal{T}$ operators, as mentioned earlier. However, $f(t)$ can be taken as a complex function satisfying (\ref{decouple}) and (\ref{susyp}). 
Taking $f(t) = f_R(t) + if_I(t)$ we arrive to the form of $f(t)$ in terms of a solution to an  Ermakov equation (see References \cite{RosasOrtiz2015,RosasOrtiz2018} for details). However, it can be shown that such an approach is equivalent to choosing 
\begin{equation}\label{efe}
    f(t) = {\dot u(t)}/{u(t)},
\end{equation}
with $u(t)$ having the form \cite{Zelaya2020}
\begin{equation}\label{ufunc}
    u(t) = c_1 u_1(t) + \left( \frac{c_2}{2} - i\frac{\lambda}{w_0} \right) u_2(t).
\end{equation}
Here, $u_j(t)$ are lineraly independent solutions of
\begin{equation}\label{epsil}
    \hat H_1 u_j(t) =\epsilon u_j(t), \quad j=1,2,
\end{equation}
with $\hat H_1$ given in \eqref{decouple} and \eqref{susyp}, where $\epsilon=-1$.
Besides, $w_0$ is the nonzero Wronskian of $u_1(t)$ and $u_2(t)$, $\lambda$ is an arbitrary real constant and $c_\ell$, $\ell = 1,2,3$ are positive numbers that must satisfy \cite{Zelaya2020} 
\begin{equation}
    c_2^2 - 4c_1c_3 ={-4\lambda^2}/{w_0^2}.
\end{equation}
If $V_1(t)$ is a real function of $t$, the functions $u_{1,2}(t)$ can also be chosen with real values. Then $\lambda$ \textit{ regulates} the imaginary part of the function $f(t)$, and the real part of the driving or coupling function in Hamiltonian (\ref{piti}). If $\lambda = 0$, the function $f(t)$ is real, and if $f(t)$ is an even function of time, the $\mathcal{PT}$-symmetry 
of the Hamiltonian (\ref{piti}) is preserved. 

In the following, a particular example of a quantum potential $V_1(t)$ satisfying (\ref{susyp}) is considered. Complex, periodic and nonperiodic, coupling functions of the form in (\ref{efe})  are obtained, for which the functions $a_{1,2}(t)$ that solve (\ref{decouple}) will be also evaluated.

\section{Example: The constant potential}\label{sec.freepotential}

\begin{figure}[t!]
\centering
\includegraphics[width=1\linewidth]{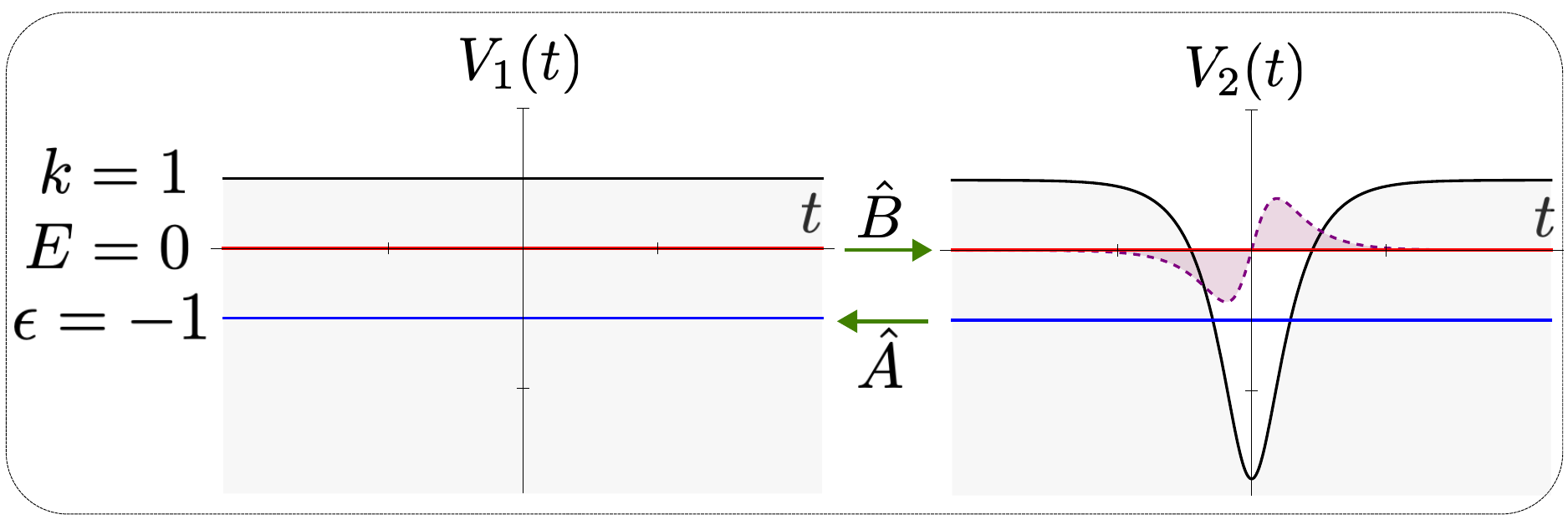}
\caption{\label{fig_susyp}  {\bf Graphical representation of the SUSY transformation.} Left panel: constant potential $V_1(t) = 1$ (black line), plus the \textit{energy} levels $E = 0$ and $\epsilon = -1$, indicated by horizontal red and blue lines, respectively. Right panel: SUSY partner $V_2(t)$ in  (\ref{PT}), for $k = 1$ and $\eta = i/4$. For the chosen values of the parameters, the SUSY partner $V_2(t)$ is a complex function of $t$. Its real (imaginary) part is given by the solid black line (dashed purple). Besides, the mapping (\ref{propto}) between solutions of the Schr\"odinger equations associated with $V_1(t)$ and $V_2(t)$, performed by the SUSY operators $\hat A$ and $\hat B$, is indicated by green horizontal arrows. The complex function $f(t)$ connecting both potentials $V_{1,2}(t)$ is given in (\ref{fv1k}).}
\end{figure}
The simplest option is $V_1(t) = k\in\mathbb R$. This choice yields the following family of solutions to  \eqref{decouple}
\begin{equation}\label{17}
    a_1(t) = \alpha_1 e^{\sqrt k t} + \alpha_2 e^{-\sqrt k t}, 
\end{equation}
with $\alpha_{1,2}\in\mathbb C$ arbitrary constants to be determined. The coupling function (\ref{susyp}) is given by
\begin{equation}\label{fv1k}
    f(t) = -\sqrt{-1-k} \tan[\sqrt{-1-k}(t + \eta)], \quad \eta\in\mathbb C,
\end{equation}
and $a_2(t)$ is found from \eqref{propto} to be
\begin{equation}\label{solutiona2}
        a_2(t) = i \alpha_1 (f(t)- \sqrt k) e^{\sqrt k t} + i \alpha_2(f(t)+ \sqrt k) e^{-\sqrt k t}.
\end{equation}
It is sufficient for our purposes to solve only one of the differential equations in (\ref{susyp}), but it is illustrative to consider the SUSY partner for a deeper understanding of the structure of the problem. For completeness, we give the expression for $V_2(t)$ from (\ref{part}) 
\begin{equation}\label{PT}
    V_2(t) = k - \frac{2(k + 1)}{\cos^2[(t + \eta)\sqrt{-1-k}]}.
\end{equation}
Figure \ref{fig_susyp} shows the potentials $V_{1,2}(t)$ found in (\ref{fv1k}) and (\ref{PT}), for some values of the parameters. Furthermore, it is schematically represented that the mapping between arbitrary \textit{eigensolutions} of $\hat H_{1,2}$ is performed using the operators
$\hat A$ and $\hat B$, as indicated in (\ref{propto}).

\subsection{Spectral analysis}\label{subsection_spectral}

\begin{figure}[t!]
\centering
\includegraphics[width=1\linewidth]{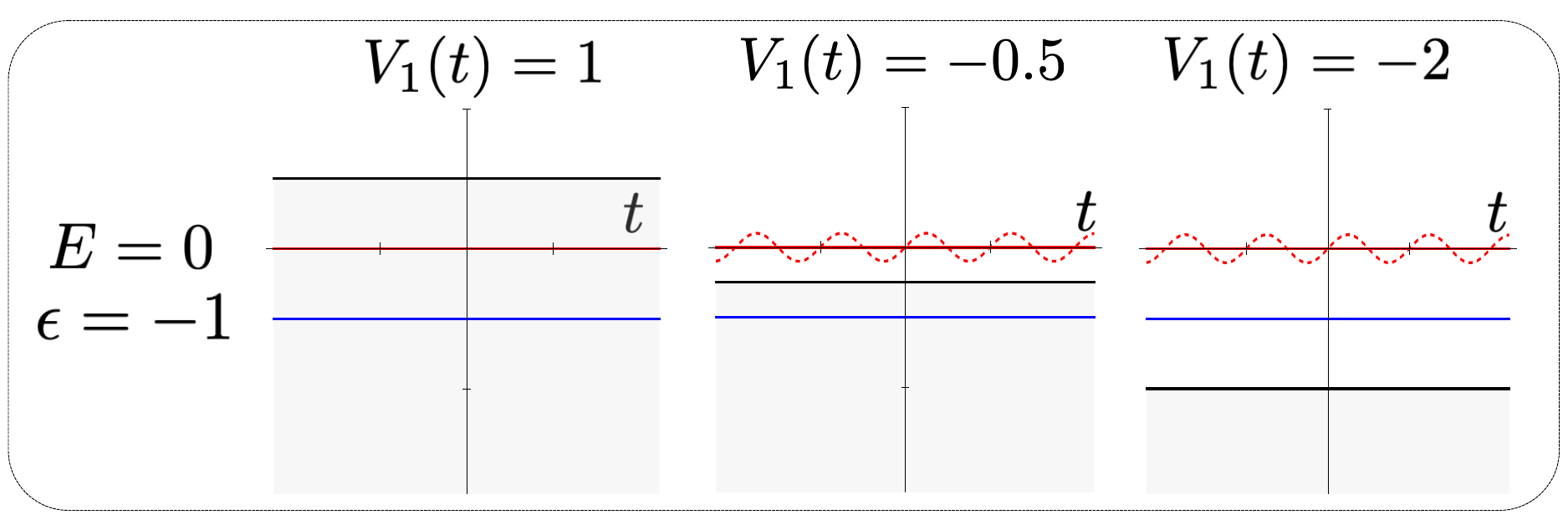}
\caption{\label{fig_spectral}  {\bf Spectral analysis of the solutions associated with the constant potential.} For fixed values of both $E = 0$ (red horizontal line) and $\epsilon = -1$ (blue horizontal line), three representative cases of $k$ (black horizontal line) have been chosen.
These are $k=1$ (left), $k = -0.5$ (middle), and $k = -2$ (right). The relative position of $k$ with respect to $E$ defines the nature of the solutions $a_{1,2}(t)$. For $E>k$ (middle and right panels), the solution $a_1(t)$ associated with $E$, is oscillatory, which is schematically represented by a dashed-red sinusoidal line. For $E < k$ (left panel), the solution $a_1(t)$ is of exponential nature. In turn, the relative position of $k$ with respect to $\epsilon$ determines the form of the superpotential $f(t)$: it is either hyperbolic ($k>\epsilon$, left and middle panels) or trigonometric ($k<\epsilon$, right panel). } 
\end{figure}

Since $a_1(t)$ is the eigenfunction of $\hat H_1$, with eigenvalue $E=0$, it seems convenient to perform a spectral analysis of the potential $V_1(t) = k$, with respect to the parameter $k$. For $k>0$ the solutions \eqref{17} and \eqref{solutiona2} are 
\begin{eqnarray}\label{a1hyp}
&&  \!\!\!  a_1(t) = \gamma_1 \cosh(\sqrt k t + \gamma_2), \\
\nonumber
 & &   \!\!\!   a_2(t) = i\gamma_1[f(t)\cosh(\sqrt k t + \gamma_2)
        - \sqrt k \sinh(\sqrt k t + \gamma_2)].
\end{eqnarray}
The constants $\gamma_{1,2}$ can be expressed in terms of the initial values $f(0)$ and $a_{1,2}(0)$ as 
$$
  \gamma_1 = \frac{a_1(0)}{\cosh\gamma_2 } , \quad 
    \gamma_2 = \mathrm
    {arctanh}\left( \frac{1}{\sqrt{|k|} }\left[ f(0) - \frac{a_2(0)}{i a_1(0)} \right] \right).
$$
On the other hand, for $k\leq 0$ (see also center and right panels in Figure \ref{fig_spectral}), the solutions $a_{1,2}(t)$ are
\begin{eqnarray}\label{a1osc}
 &&  \!\!\!\!\!\!  a_1(t) = \beta_1\cos(\sqrt{|k|}t + \beta_2),
\\
  &&   \!\!\!\!\!\!   a_2(t) = i\beta_1[f(t)\cos(\sqrt{|k|} t + \beta_2)
        + \sqrt{|k|} \sin(\sqrt{|k|} t + \beta_2)], \nonumber
\end{eqnarray}
with the constants $\beta_{1,2}\in\mathbb C$ given as 
$$
    \beta_1 = \frac{a_1(0)}{\cos\beta_2 } ,
\quad    \beta_2 = \arctan\left( \frac{1}{\sqrt{|k|} }\left[ \frac{a_2(0)}{i a_1(0)} - f(0) \right] \right).
$$

Thus, depending on $k$ the solutions $a_1(t)$ and $a_2(t)$ are either exponential ($k>0$) or oscillatory ($k<0$) over time. Also, according to (\ref{efe})-(\ref{epsil}), $f(t)$ will be hyperbolic ($k>-1$) or trigonometric ($k<-1$).

\subsection{Population inversion and purely imaginary drivings (real superpotentials)}
A dynamical variable of interest in a two-level system is the population inversion 
\begin{equation}\label{inv}
    W(t) = \langle\phi|\hat\sigma_z|\phi\rangle = |c_1(t)|^2 - |c_2(t)|^2,
\end{equation}
which gives the difference between the probabilities of finding the qubit in the excited and the ground states, i.e. $W(t) = p_{ee}(t) - p_{gg}(t)$, where
\begin{equation}
    |\phi\rangle=\left(\begin{array}{cc}
     c_1(t)  \\
     c_2(t) 
\end{array} \right) = \frac{1}{\sqrt 2}\left(\begin{array}{cc}
     1 & -1 \\
     1 & 1 
\end{array} \right) \left(\begin{array}{cc}
     a_1(t)  \\
     a_2(t) 
\end{array} \right).
\end{equation}
It can be shown that
\begin{equation}
    W(t) = -2 [a_1^R(t) a_2^R(t) + a_1^I(t) a_2^I(t)],
\end{equation}
where the superscripts denote the real ($R$) or imaginary ($I$) parts of the functions $a_{1,2}(t)$.
Taking into account the first expression in \eqref{decouple} and (\ref{propto}), it can be proved that
\begin{equation}\label{const_inv}
    \partial_t W(t) \propto f_I(t), \quad\textrm{for all $t$}.
\end{equation}
This means that a purely imaginary coupling in the Hamiltonian in (\ref{piti}) does not produce population transitions, in accordance with what is stated at the end of Section II in \cite{Gong2015}. However, this does not occur when $f(t)$ has a nontrivial imaginary part, giving rise to a real part of the coupling function in the Hamiltonian (\ref{piti}).

In the following, we present numerical examples of complex time-dependent coupling functions, for which solutions of (\ref{piti}) are known, and which give rise to controlled transitions in the qubit state.

\section{Numerical example: Complex superpotential}\label{sec.numerical}

Below are some numerical experiments to elucidate the applicability of the results obtained in the previous sections.
All analytical solutions were tested by direct comparison with the numerical solution of equation (\ref{piti}), which was solved by implementing a fourth-order Runge-Kutta numerical method, where the step was appropriately fixed after a convergence test.
The qubit splitting energy $\Delta$ sets the time scale, via the change of variable indicated above (\ref{dimensionless}).
The remaining parameters are also given in units of $\Delta$ ($k$) and $\Delta^{-1}$ ($\eta$).

From to the general form of the function $f(t)$ in (\ref{fv1k}), we get
\begin{equation}\label{f1f2f3}
    f(t) = \left\{
    \begin{matrix}
        \kappa \tanh[\kappa (t + \eta)], & k > 0,\\
        \nu \tanh[\nu (t + \eta)], & -1 \leq k \leq 0,\\
        -\varepsilon \tan[\varepsilon (t + \eta)], & k < -1,
    \end{matrix}
    \right.
\end{equation}
where
\begin{equation}\label{cnstnt}
        \kappa = \sqrt{1 + k},\quad
        \nu = \sqrt{1 - |k|},\quad
        \varepsilon  = \sqrt{|k| - 1}.\\
\end{equation}
Here, a complex \textit{displacement} $\eta = \theta + i\varphi$, $\theta,\varphi\in\mathbb R$, is considered. In addition, the solutions $a_{1,2}(t)$ are chosen accordingly to the considered value (interval) of $k$, as given in Section~\ref{subsection_spectral}.

Note that from (\ref{f1f2f3}) and (\ref{cnstnt}), that the nondriven case $f(t) = 0$ can be obtained in the limit $k\to -1$.

\subsection{Decaying solutions}
For $k>0$, a bounded decaying solution can be obtained in the limit $a_2(0)\to ia_1(0)[f(0) + \sqrt k]$. In addition, to fulfill $|a_1(0)|^2 + |a_2(0)|^2 = 1$ we set
\begin{equation}\label{a10}
    a_1(0) = \left( 1 + \left|f(0) + \sqrt k \right|^2 \right)^{-1/2}.
\end{equation}
The complex coupling function
\begin{equation}\label{f1}
    f(t) = \kappa\tanh[\kappa(t + \theta + i\varphi)],\qquad \kappa = \sqrt{1 + k},
\end{equation}
can be obtained from (\ref{efe}) and (\ref{ufunc}) by setting 
\begin{equation}\label{changes}
    \begin{split}
        c_1 &=\cos(\kappa\varphi),\qquad\
        c_2 = 0,\qquad
        \lambda = -\kappa\sin(\kappa\varphi),\\
        u_1(t) &= \cosh(\kappa t),\qquad
        u_2(t) = \sinh(\kappa t),\\
      &  t\mapsto t + \theta, \qquad\quad \theta\in\mathbb R. 
    \end{split}
\end{equation}

\begin{figure}[t!]
\centering
\includegraphics[width=1\linewidth]{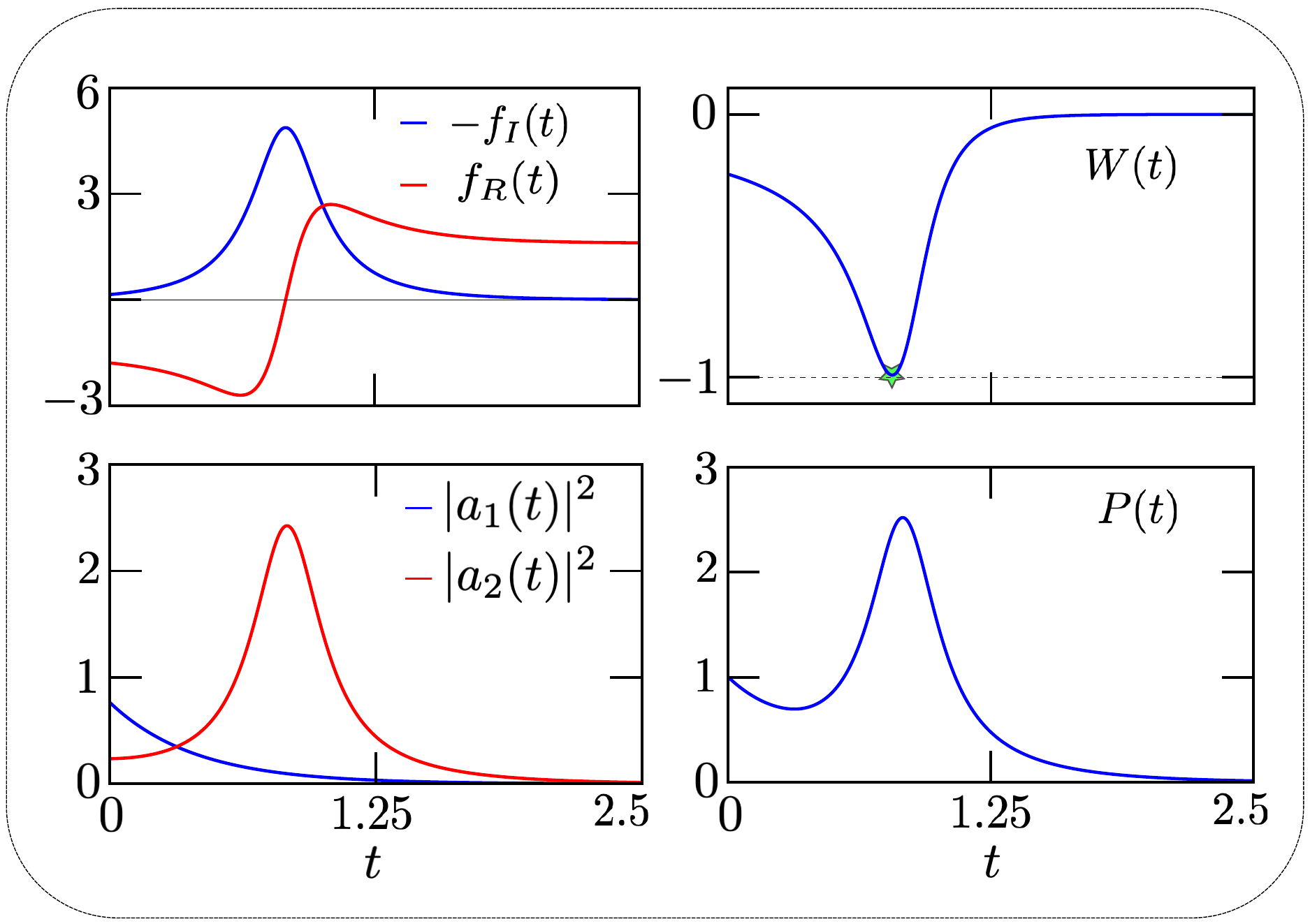}
\caption{\label{fig_decaying}  {\bf Non-Hermitian dynamics of decaying solutions.} 
Top left:  Complex driving defined by (\ref{f1}). Top right: The atomic inversion (\ref{inv}) shows a transition from a point in the lower hemisphere ($W(0)<0$) of the Bloch sphere to the ground state $|g\rangle$ ($W(t_1)=-1$).
This value is marked with a green star. Bottom left: square moduli of the solutions $a_{1,2}(t)$. The exponential decay of $a_1(t)$ can be appreciated (blue).
Bottom right: The quantity $P(t)= |c_1(t)|^2 + |c_2(t)|^2$ is shown as evidence that unitarity is not preserved in the non-Hermitian system defined by (\ref{piti}). For all figures the parameter values are $k = 1.566$, $\theta = -0.83$ and $\varphi = 3.14$. }
\end{figure}

Figure \ref{fig_decaying} shows the atomic inversion $W(t)$ (top right panel) for the complex driving $w(t)/\Delta =if(t)$ (top left panel) defined by (\ref{f1}), for some values of the parameters.
It can be seen that $W(0) < 0$, which means that the initial state of the qubit system lies somewhere in the lower hemisphere of the Bloch sphere (cf. Figure~\ref{fig_gainloss}A).
However, at some instant $t_1>0$, we have $W(t_1) = -1$, which means that the state of the qubit is $|\phi(t_1)\rangle = |g\rangle$ with absolute certainty.
This value of $W$ is marked with a green star in the upper right panel of Figure~\ref{fig_decaying}.

On the other hand, the lower panels of Figure \ref{fig_decaying} exemplify the nonconservative nature of the dynamics due to the non-Hermitian Hamiltonian (\ref{piti}). The lower left panel shows the square moduli of $|a_{1,2}(t)|^2$ and here you can see the exponential decay of $a_1(t)$ (blue line). However, we observe that $|a_2(t)|^2>1$ (red line) for a certain time interval; we attribute this to the non-Hermitian (nonconservative) nature of Hamiltonian (\ref{piti}).
Furthermore, the quantity defined as $P(t)=|c_1(t)|^2 + |c_2(t)|^2$ is presented in the lower right panel of Figure \ref{fig_decaying}, where the behavior $P(t)>1$ over a time interval is also shown, though $P(0) = 1$. This behavior is naturally inherited from $|a_2(t)|^2$.

The parameters chosen in Figure \ref{fig_decaying} are $k = 1.566$, $\theta = -0.83$, $\varphi = 3.14$. In turn, the initial condition is set so that $a_1(0)$ is taken as in (\ref{a10}) and $a_2(0) \to ia_1(0) [ f(0) + \sqrt k ]$, as mentioned above. 
If $a_2(0)$ is not chosen as previously mentioned, exponentially increasing (unbounded) solutions $c_{1,2}(t)$ are obtained. 
Furthermore, the Hamiltonian (\ref{piti}) for the coupling defined by (\ref{f1}) is $\mathcal{PT}$-symmetric, not in the sense described in Section~\ref{sec.model}, but by defining the parity operation $\mathcal{P}$ as $x\mapsto -x$, where $x = t +\theta$, while the time reversal operation $\mathcal{T}$ remains the usual complex conjugation. 

\subsection{Oscillatory solutions}

We now analyze the case $k \leq 0$. However, we divide the problem according to the form of the superpotential $f(t)$, i.e. hyperbolic ($-1 \leq k \leq 0$) or trigonometric ($k < -1$). As above, the presented cases are $\mathcal{PT}$-symmetric, where the parity operation $\mathcal{P}$ performs $x\mapsto -x$ and the time-reversal operation $\mathcal{T}$ performs complex conjugation.

\subsubsection{Hyperbolic superpotential}\label{sec_hyper}

For the case $-1 \leq k \leq 0$, the complex function
\begin{equation}\label{f2}
    f(t) = \nu\tanh[\nu(t + \theta + i\varphi)],\qquad \nu = \sqrt{1 - |k|},
\end{equation}
similar to (\ref{f1}), can be obtained by the changes given in (\ref{changes}), after replacing $\kappa\mapsto\nu$.

\begin{figure}[!]
\centering
\includegraphics[width=1\linewidth]{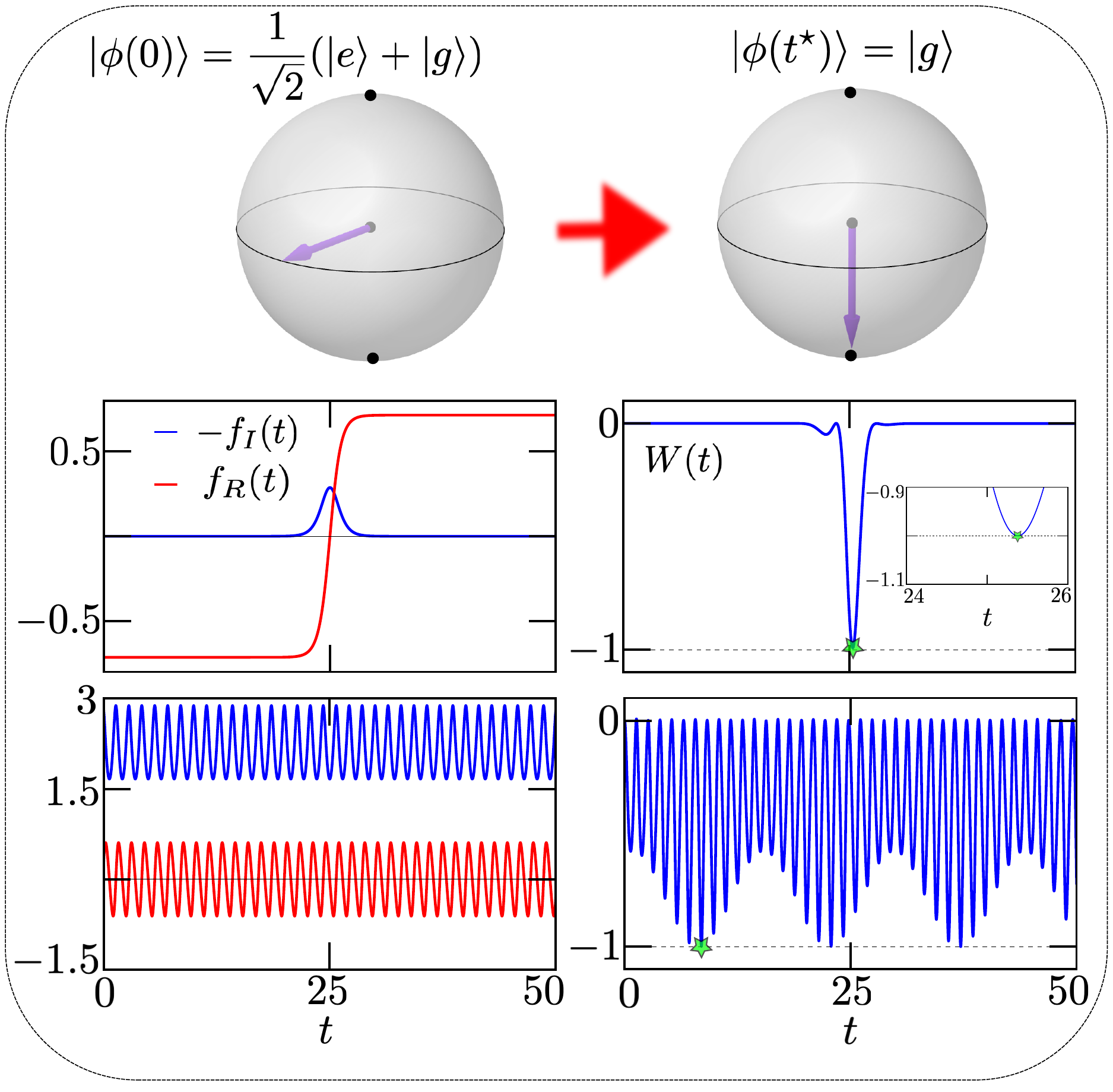}
\caption{\label{fig_oscillatory}  {\bf Transitions in the qubit state for the case of oscillatory solutions.}  
Top figure: Transition between two particular states of the qubit. From $|\phi(0)\rangle = \frac{1}{\sqrt 2}(|e\rangle + |g\rangle)$ (top left) to $|\phi(t^\star) = |g\rangle$ (top right), for a fixed instant of time $t = t^\star$. 
Middle panels: hyperbolic driving $w(t)/\Delta =if(t)$ (middle left), with $f(t)$ as given in (\ref{f2}), and the  atomic inversion $W(t)$ (middle right). 
Bottom panels: trigonometric driving $w(t)/\Delta =if(t)$ (bottom left), with $f(t)$ as given in (\ref{f3}), and the atomic inversion $W(t)$ (bottom right). 
The time $t=t^\star$ at which $|\phi(t^\star) = |g\rangle$ is different for the hyperbolic (middle right) and trigonometric (bottom right) drivings. However, in both cases, the value $W = -1$ is marked with a green star. The parameter values for the middle panels are: $k=-0.49$, $\theta = -25$, $\varphi=-0.54$, with initial condition $c_1(0) = c_2(0) = 1/\sqrt 2$. Those for the lower panels are: $k=-5.8$, $\theta = -25$ and $\varphi=0.455$, for the initial condition $c_1(0) = -c_2(0) = - 1/\sqrt 2$. } 
\end{figure}

Figure \ref{fig_oscillatory} shows that it is possible to generate transitions in the qubit, in the case of complex driving functions in the Hamiltonian (\ref{piti}). 
Specifically, from the initial state $|\phi(0)\rangle = 1/\sqrt 2 (|e\rangle + |g\rangle)$ to the state $|\phi(t^\star)\rangle =|g\rangle$ in a time interval $t^\star$ (top row in Figure \ref{fig_oscillatory}). 
In particular, the middle row of Figure \ref{fig_oscillatory} shows the case $-1 \leq k \leq 0$: the complex driving $w(t)/\Delta =if(t)$ (middle left) and the atomic inversion $W(t)$ (middle right), according to (\ref{f2}) and (\ref{inv}), respectively. 
The parameters considered for the panels in the middle row are: $k=-0.49$, $\theta = -25$, $\varphi=-0.54$, with the initial condition $c_1(0) = c_2(0) = 1/\sqrt 2$.

The atomic inversion (middle right panel) shows peculiar behavior. It remains unchanged and equal to zero for a relatively long time interval, corresponding to $f_I = 0$, in agreement with (\ref{const_inv}). 
Then, just before $t=25$, it exhibits small oscillations and \textit{suddenly} goes to $W(t) = -1$, at $t = t^\star$. This value of $W(t^\star)$ is marked by a green star in the middle right panel. 
In turn, the value $W(t^\star) = -1$ represents the change from the initial state $|\phi(0)\rangle = 1/\sqrt 2 (|e\rangle + |g\rangle)$ to the state $|\phi(t^\star)\rangle =|g\rangle$.
It is remarkable that, in the presence of complex driving, as defined in (\ref{piti}), such atomic transitions can still be achieved. Note also that at $t=t^\star$ the imaginary part of the driving or coupling $w(t)$, i.e. $f_R$, is different from zero. 
In fact, $f_R(t=25) = 0$ for the chosen parameter values, although $t^\star\neq 25$, as can be seen in the inset of the middle right panel of Figure \ref{fig_oscillatory}. 
Thus, the sharp peak shown by the atomic inversion, and defining the qubit transition, in the case $-1 \leq k \leq 0$ can be judiciously stimulated by the complex driving defined by (\ref{f2}). 
The transition between $|\phi(0)\rangle = |e\rangle$ and $|\phi(t^\star) = |g\rangle$ (or vice versa), for a given $t^\star$, can be promoted by adjusting the interaction parameters.

\begin{figure}[b!]
\centering
\includegraphics[width=1\linewidth]{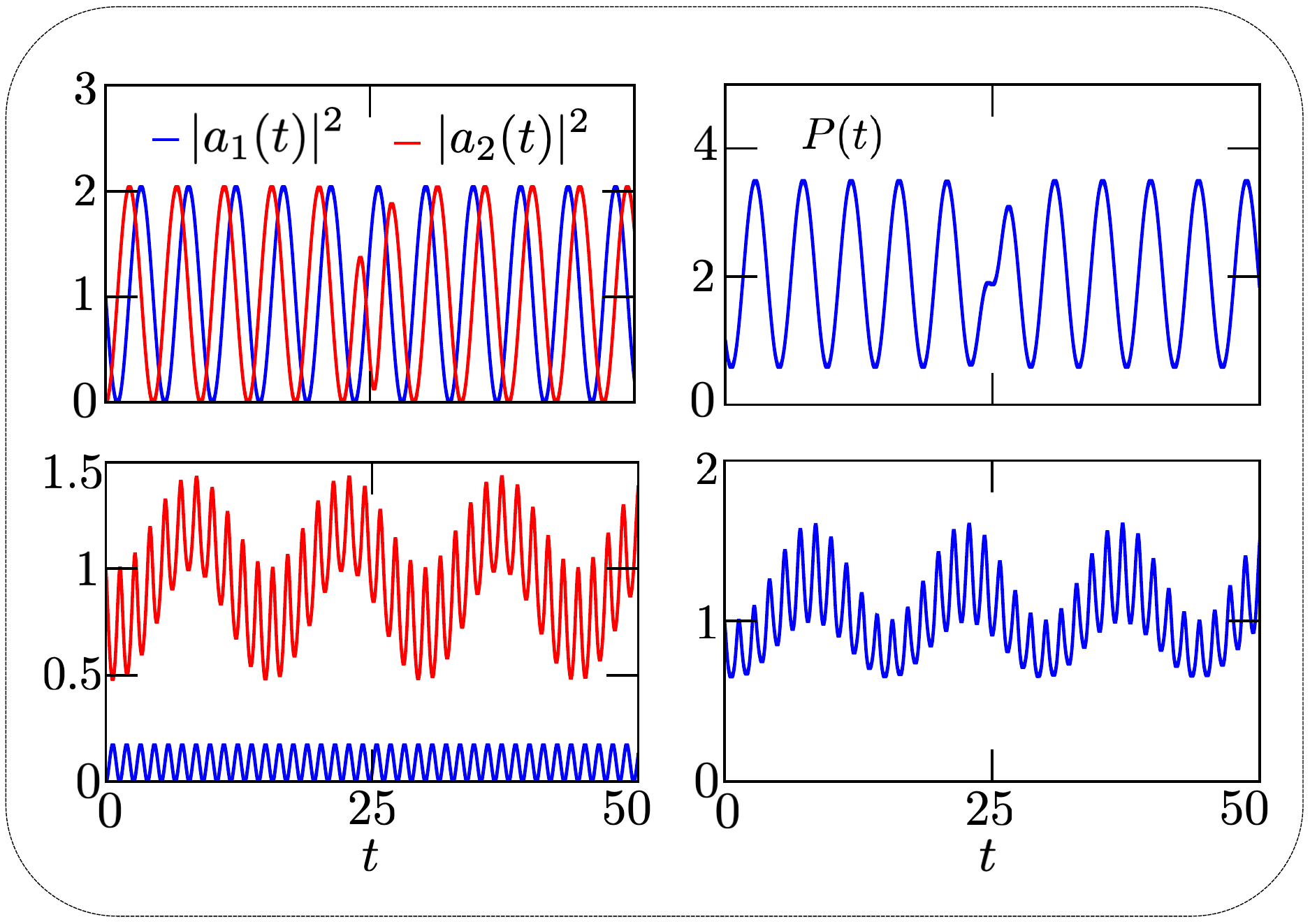}
\caption{\label{fig_oscillatory_aux}  {\bf Non-Hermitian dynamics of oscillatory solutions.} Modulus squared of the solutions $a_{1,2}(t)$ (left column) and the  $P(t)$ function (right column), corresponding to the middle (top row) and lower (bottom row) panels of Figure \ref{fig_oscillatory}. The oscillatory behavior of $a_{1,2}(t)$ can be appreciated (left column). Furthermore, the nonconservative dynamics associated with the Hamiltonian (\ref{piti}) can be recognized. This can also be inferred from the function $P(t)$  (right column).} 
\end{figure}

The top row of Figure \ref{fig_oscillatory_aux} shows the square moduli of the solutions $a_{1,2}(t)$ (top left) and their sum $P(t)$ (top right) for the hyperbolic driving case (\ref{f2}). Both the oscillatory behavior of $a_{1,2}(t)$ and the nonconservative nature associated with the Hamiltonian (\ref{piti}) can be noticed, via $|a_{1,2}(t)|^2$ (top left) and the function $P(t)$ (top right). 
In particular, it is intriguing that although both $|a_{1,2}(t)|^2$ and $P(t)$ exhibit oscillatory behavior, such oscillations are not present in $W(t)$ (middle right panel of Figure \ref{fig_oscillatory}). Also, note that around $t = 25$ both $|a_{1,2}(t)|^2$ and $P(t)$ change their behavior, 
giving rise to the qubit state transition shown in the middle right panel of Figure \ref{fig_oscillatory}.

\subsubsection{Trigonometric superpotential}

If $k < -1$, the complex function
\begin{equation}\label{f3}
    f(t) = - \varepsilon\tan[\varepsilon(t + \theta + i\varphi)],\qquad \varepsilon=\sqrt{|k|-1},
\end{equation}
can be obtained from (\ref{efe}) and (\ref{ufunc}), by setting 
\begin{eqnarray}
  &&      c_1 =\cosh(\varepsilon\varphi),\quad c_2 = 0,\quad
        \lambda = \varepsilon\sinh(\varepsilon\varphi),\\
  &&      u_1(t) = \cos(\varepsilon t),\quad
        u_2(t) = \sin(\varepsilon t),\\
   &&   t\mapsto t + \theta.
\end{eqnarray}
Thus, the trigonometric tangent becomes nonsingular in $\mathbb R$, because $\varphi\neq 0$.

The bottom row of Fig. \ref{fig_oscillatory} shows the complex driving function $w(t)/\Delta =if(t)$ (bottom left) and the atomic inversion $W(t)$ (bottom right), according to (\ref{f3}) and (\ref{inv}), respectively, for the case $k < -1$. We chose the parameter values $k=-5.8$, $\theta = -25$ and $\varphi=0.455$, for the initial condition $c_1(0) = -c_2(0) = - 1/\sqrt 2$.
The oscillatory nature of both the complex driving function and atomic inversion can be appreciated. Specifically, the atomic inversion $W(t)$ exhibits oscillations that increase in amplitude. 
As in the previous case, at a given instant $t=t^\star$, different from the one that appears in the case $-1 \leq k \leq 0$ and also marked with a green star in the bottom left panel of Fig.~\ref{fig_oscillatory}, it is found that $W(t^\star) = -1$. Thus, the oscillatory nature of $c_{1,2}(t)$ produces the transition from $|\phi(0)\rangle = 1/\sqrt 2 (|e\rangle + |g\rangle)$ to $|\phi(t^\star)\rangle =|g\rangle$. Unlike the case $-1 \leq k \leq 0$, where the change appears \textit{suddenly}, now such a transition is achieved after several periods.
Furthermore, $W(t)$ shows a general periodic evolution, reaching the value $W(t) = -1$ not only once, as in the previous case, but periodically in time; however, only the first value $W = -1$ has been marked with a green star. Therefore, it refers to periodic transitions between states $|e\rangle + |g\rangle$ and $|g\rangle$.

The bottom row of Fig. \ref{fig_oscillatory_aux} shows the square moduli $|a_{1,2}(t)|^2$ (bottom left) and the function $P(t)$ (bottom right) for the case $k<-1$. As above, one can see the periodic structure of both $|a_{1,2}(t)|^2$ and $P(t)$, which in turn give rise to the periodic qubit transitions predicted by $W(t)$ in the bottom right panel of Fig. \ref{fig_oscillatory}.
Also, the nonconservative nature of (\ref{piti}) can be appreciated, particularly from $P(t)$ (bottom right panel in Fig. \ref{fig_oscillatory_aux}).

In summary, we observe that the oscillatory solutions $a_{1,2}(t)$ are useful to achieve transitions between the states $|\phi(0)\rangle = 1/\sqrt 2 (|e\rangle + |g\rangle)$ and $|\phi(t^\star)\rangle =|g\rangle$, for a given time interval $t^\star$, for both the complex hyperbolic (\ref{f2}) and for the trigonometric (\ref{f3}) drivings. These, in turn, have different characteristics depending on the intrinsic shape of the superpotential $f(t)$.

\section{Proposed experimental observations}\label{sec.experiment}

There are several experimental platforms on which the non-Hermitian Hamiltonian (\ref{piti}) can be easily implemented. Here, we briefly review the actual experimental possibilities.
First, recall that driven two-level systems usually appear in the context of nuclear magnetic resonance (NMR) \cite{Bloch1946,Silver1984PRA,Silver1984JMR,Silver1985,Giscard2020}). Numerous driven Hermitian systems have been considered in the existing literature (see, e.g., \cite{Silver1985,Martinez2023,Enriquez2018}). 
In the non-Hermitian regime, several $\mathcal{PT}$-symmetric systems have been studied \cite{Bender2007,Moiseyev2011}, and in particular the non-Hermitian system defined by (\ref{piti}) can also be interpreted in the context of NMR. To see this, compare the Hamiltonian (\ref{piti}) with the one given in equation (27) of Ref. \cite{Martinez2023}
\begin{equation}
    \hat H_{ \rm{Ref} }(t) = \frac{1}{2}\Omega(t)\hat T_+ + \delta\omega\hat T_c + \frac{1}{2}\Omega^*(t)\hat T_-,
\end{equation}
where the operators $\hat T_\pm$, $\hat T_c$ are $\hat\sigma_\pm$ and $\hat\sigma_z/2$, and the star denotes complex conjugation. In our case, (\ref{piti}) has the form
\begin{equation}
    \hat H(t) = \frac{1}{2}\mathcal O(t)\hat T_+ + \delta\omega\hat T_c + \frac{1}{2}\mathcal O(t)\hat T_-,
\end{equation}
with $\mathcal O(t) = -2ig(t)$ and $\delta\omega = \Delta$. This leads to a slight modification of the Ricatti equation (26) in \cite{Martinez2023}:
\begin{equation}
    \dot f - \frac{i}{2} \Omega^*(t) f^2 + i\Delta\omega f + \frac{i}{2}\Omega(t) = 0.
\end{equation}
Instead, we get
\begin{equation}
    \dot F - \frac{i}{2} \mathcal O(t) F^2 + i\Delta\omega F + \frac{i}{2}\mathcal O(t) = 0.
\end{equation}
This modifies the structure of the Bloch equations (24) in \cite{Martinez2023}, leading to a non-Hermitian version of the usual NMR scheme. Indeed, it is this slight modification that is responsible for the non-Hermitian nature of (\ref{piti}) and hence the gain/loss of the current system. Moreover, according to \cite{Moiseyev2011}, non-Hermitian Hamiltonians like (\ref{piti}) model the interactions between metastable (resonant) states of the two-level system.

On the other hand, the non-Hermitian Hamiltonian (\ref{piti}) can be directly implemented in the context of coupled waveguides, i.e., as non-Hermitian (integrated gain/loss \cite{Ruter2010}) directional couplers (i.e., within the framework of the strong-binding model), where both the refractive index and the gain/loss profile depend on the longitudinal direction of propagation. The reader is referred, for example, to (1) in \cite{Liu2020}, which can be compared with (\ref{piti}) in the present manuscript (see also \cite{Luo2013}). Besides, note the similarities between Hamiltonian (1) in \cite{Hassan2017} and the one in \eqref{pete}. This implementation is feasible with the technology currently available \cite{Ruter2010,Hassan2017}.

\section{Conclusions}\label{sec.conclusions}

The SUSY technique allows for the straightforward construction of complex drivings for the semiclassical Rabi system defined by (\ref{piti}), as well as for solutions of the time-dependent Schr\"odinger equation. 
Therefore, the known SUSY systems available in the literature \cite{Cooper1995,fernandez1999,Mielnik2000,Samsonov2004,Correa2007,miri2013,Correa2015,CruzyCruz2020,Garcia2023} are suitable to be used to produce complex driving functions $w = ig(t)$ in (\ref{piti}). 
This, in turn, will increase the number of time-dependent, exactly solvable (analytically solvable)  non-Hermitian problems, specifically those of the type (\ref{piti}). 
It is important to emphasize that the solution of the Schr\"odinger equation (\ref{piti}) for the coupling function(s) defined by (\ref{f1f2f3}) is not trivial at all, since the Hamiltonian given in (\ref{piti}) for such choices of $g(t)$ is non-Hermitian and time-dependent. However, as shown, the SUSY approach presented in this manuscript turns such a task into a completely tractable analytical problem.

In particular, the simplest case of a constant potential $V_1$ allows the construction of complex driving functions (\ref{f1f2f3}), which produce transitions in the qubit state. 
Two types of controlled transitions were obtained: periodic and nonperiodic, depending on the nature of the coupling or driving function.
In the frame of Ref.~\cite{Gong2015}, it is essential to mention that the present results prove that not only a periodic impulse stabilizes the dynamics of a non-Hermitian system, but also a nonperiodic modulation function $g(t)$, as shown in section \ref{sec_hyper}.
Moreover, this is just the beginning, as there are still many SUSY pairs to be analyzed \cite{Cooper1995,fernandez1999,Mielnik2000,Samsonov2004,Correa2007,miri2013,Correa2015,CruzyCruz2020,Garcia2023}. 
Specifically, a quantum potential $V_1$ with a point-like spectrum (with bound states) could expand the richness of the dynamics of the time-dependent non-Hermitian problem. For example, the Scarf II-type potential seems a good candidate for testing the current formalism \cite{Bocanegra2022}, which is proposed as future work. Furthermore, the present analysis elucidates the importance of the theoretical study of the SUSY partners of quantum potentials, particularly the free particle potential.

Certainly, non-Hermitian Hamiltonians like (\ref{piti}) are of essential importance, because they describe physical interactions in the context of NMR (\cite{Bloch1946,Silver1984PRA,Silver1984JMR,Silver1985,Martinez2023,Enriquez2018}).
In particular, they model interactions between metastable states (resonances), e.g., an atom interacting with combined AC and DC fields \cite{Moiseyev2011}. 
Furthermore, the non-Hermitian Hamiltonian (\ref{piti}) can be simulated by a pair of electromagnetic waveguides coupled via evanescent fields (as a non-Hermitian directional coupler), once the refractive indices and the gain/loss profile are allowed to depend on the propagation coordinate \cite{Liu2020,Luo2013}.

Finally, it is worth mentioning that the presented results are in agreement with previously published conclusions \cite{Gong2015,Martinez2023,Enriquez2018} and give a deep, complementary insight into the dynamics and overall behavior of non-Hermitian systems that in turn are important in current major scenarios as shortcuts to adiabaticity \cite{Bender2007,Ibanez2011,Zheng2013,Torosov2013}, phase transitions \cite{Pires2021}, quantum chaos \cite{Cornelius2022},  to mention a few.

\section*{Acknowledgments}

We thank C. M. Bender, D. Braak, A.N. Ikot, J. Negro, and A. Perez-Leija for fruitful discussions and comments. This work was supported by Spanish MCIN with funding from European Union Next Generation EU (PRTRC17.I1) and Consejeria de Educacion from Junta de Castilla y Leon through the QCAYLE project, as well as Grant No. PID2023-148409NB-I00 MTM funded by AEI/10.13039/501100011033, and RED2022-134301-T. The financial support of the Department of Education of the Junta de Castilla y Leon, and the FEDER Funds is also appreciated (CLU-2023-1-05).

\bibliography{My_Bib_SUSY_JC_LMN}

\begin{thebibliography}{62}%
\makeatletter
\providecommand \@ifxundefined [1]{%
 \@ifx{#1\undefined}
}%
\providecommand \@ifnum [1]{%
 \ifnum #1\expandafter \@firstoftwo
 \else \expandafter \@secondoftwo
 \fi
}%
\providecommand \@ifx [1]{%
 \ifx #1\expandafter \@firstoftwo
 \else \expandafter \@secondoftwo
 \fi
}%
\providecommand \natexlab [1]{#1}%
\providecommand \enquote  [1]{``#1''}%
\providecommand \bibnamefont  [1]{#1}%
\providecommand \bibfnamefont [1]{#1}%
\providecommand \citenamefont [1]{#1}%
\providecommand \href@noop [0]{\@secondoftwo}%
\providecommand \href [0]{\begingroup \@sanitize@url \@href}%
\providecommand \@href[1]{\@@startlink{#1}\@@href}%
\providecommand \@@href[1]{\endgroup#1\@@endlink}%
\providecommand \@sanitize@url [0]{\catcode `\\12\catcode `\$12\catcode
  `\&12\catcode `\#12\catcode `\^12\catcode `\_12\catcode `\%12\relax}%
\providecommand \@@startlink[1]{}%
\providecommand \@@endlink[0]{}%
\providecommand \url  [0]{\begingroup\@sanitize@url \@url }%
\providecommand \@url [1]{\endgroup\@href {#1}{\urlprefix }}%
\providecommand \urlprefix  [0]{URL }%
\providecommand \Eprint [0]{\href }%
\providecommand \doibase [0]{http://dx.doi.org/}%
\providecommand \selectlanguage [0]{\@gobble}%
\providecommand \bibinfo  [0]{\@secondoftwo}%
\providecommand \bibfield  [0]{\@secondoftwo}%
\providecommand \translation [1]{[#1]}%
\providecommand \BibitemOpen [0]{}%
\providecommand \bibitemStop [0]{}%
\providecommand \bibitemNoStop [0]{.\EOS\space}%
\providecommand \EOS [0]{\spacefactor3000\relax}%
\providecommand \BibitemShut  [1]{\csname bibitem#1\endcsname}%
\let\auto@bib@innerbib\@empty
\bibitem [{\citenamefont {Rabi}(1936)}]{Rabi1936}%
  \BibitemOpen
  \bibfield  {author} {\bibinfo {author} {\bibfnamefont {I.~I.}\ \bibnamefont
  {Rabi}},\ }\href {\doibase 10.1103/PhysRev.49.324} {\bibfield  {journal}
  {\bibinfo  {journal} {Phys. Rev.}\ }\textbf {\bibinfo {volume} {49}},\
  \bibinfo {pages} {324} (\bibinfo {year} {1936})}\BibitemShut {NoStop}%
\bibitem [{\citenamefont {Rabi}(1937)}]{Rabi1937}%
  \BibitemOpen
  \bibfield  {author} {\bibinfo {author} {\bibfnamefont {I.~I.}\ \bibnamefont
  {Rabi}},\ }\href {\doibase 10.1103/PhysRev.51.652} {\bibfield  {journal}
  {\bibinfo  {journal} {Phys. Rev.}\ }\textbf {\bibinfo {volume} {51}},\
  \bibinfo {pages} {652} (\bibinfo {year} {1937})}\BibitemShut {NoStop}%
\bibitem [{\citenamefont {Braak}\ \emph {et~al.}(2016)\citenamefont {Braak},
  \citenamefont {Chen}, \citenamefont {Batchelor},\ and\ \citenamefont
  {Solano}}]{Braak2016}%
  \BibitemOpen
  \bibfield  {author} {\bibinfo {author} {\bibfnamefont {D.}~\bibnamefont
  {Braak}}, \bibinfo {author} {\bibfnamefont {Q.-H.}\ \bibnamefont {Chen}},
  \bibinfo {author} {\bibfnamefont {M.~T.}\ \bibnamefont {Batchelor}}, \ and\
  \bibinfo {author} {\bibfnamefont {E.}~\bibnamefont {Solano}},\ }\href
  {\doibase 10.1088/1751-8113/49/30/300301} {\bibfield  {journal} {\bibinfo
  {journal} {J. Phys. A: Math. Theor.}\ }\textbf {\bibinfo {volume} {49}},\
  \bibinfo {pages} {300301} (\bibinfo {year} {2016})}\BibitemShut {NoStop}%
\bibitem [{\citenamefont {Lefebvre}\ \emph {et~al.}(2009)\citenamefont
  {Lefebvre}, \citenamefont {Atabek}, \citenamefont {\ifmmode~\check{S}\else
  \v{S}\fi{}indelka},\ and\ \citenamefont {Moiseyev}}]{Moiseyev2009}%
  \BibitemOpen
  \bibfield  {author} {\bibinfo {author} {\bibfnamefont {R.}~\bibnamefont
  {Lefebvre}}, \bibinfo {author} {\bibfnamefont {O.}~\bibnamefont {Atabek}},
  \bibinfo {author} {\bibfnamefont {M.}~\bibnamefont {\ifmmode~\check{S}\else
  \v{S}\fi{}indelka}}, \ and\ \bibinfo {author} {\bibfnamefont
  {N.}~\bibnamefont {Moiseyev}},\ }\href {\doibase
  10.1103/PhysRevLett.103.123003} {\bibfield  {journal} {\bibinfo  {journal}
  {Phys. Rev. Lett.}\ }\textbf {\bibinfo {volume} {103}},\ \bibinfo {pages}
  {123003} (\bibinfo {year} {2009})}\BibitemShut {NoStop}%
\bibitem [{\citenamefont {Moiseyev}(2011)}]{Moiseyev2011}%
  \BibitemOpen
  \bibfield  {author} {\bibinfo {author} {\bibfnamefont {N.}~\bibnamefont
  {Moiseyev}},\ }\href {\doibase 10.1103/PhysRevA.83.052125} {\bibfield
  {journal} {\bibinfo  {journal} {Phys. Rev. A}\ }\textbf {\bibinfo {volume}
  {83}},\ \bibinfo {pages} {052125} (\bibinfo {year} {2011})}\BibitemShut
  {NoStop}%
\bibitem [{\citenamefont {Lee}\ and\ \citenamefont {Joglekar}(2015)}]{Lee2015}%
  \BibitemOpen
  \bibfield  {author} {\bibinfo {author} {\bibfnamefont {T.~E.}\ \bibnamefont
  {Lee}}\ and\ \bibinfo {author} {\bibfnamefont {Y.~N.}\ \bibnamefont
  {Joglekar}},\ }\href {\doibase 10.1103/PhysRevA.92.042103} {\bibfield
  {journal} {\bibinfo  {journal} {Phys. Rev. A}\ }\textbf {\bibinfo {volume}
  {92}},\ \bibinfo {pages} {042103} (\bibinfo {year} {2015})}\BibitemShut
  {NoStop}%
\bibitem [{\citenamefont {Castaños}(2019)}]{Castanos2019}%
  \BibitemOpen
  \bibfield  {author} {\bibinfo {author} {\bibfnamefont {L.}~\bibnamefont
  {Castaños}},\ }\href {\doibase 10.1016/j.optcom.2018.08.046} {\bibfield
  {journal} {\bibinfo  {journal} {Opt. Commun.}\ }\textbf {\bibinfo {volume}
  {430}},\ \bibinfo {pages} {176–188} (\bibinfo {year} {2019})}\BibitemShut
  {NoStop}%
\bibitem [{\citenamefont {Gangopadhyay}\ \emph {et~al.}(2010)\citenamefont
  {Gangopadhyay}, \citenamefont {Dzero},\ and\ \citenamefont
  {Galitski}}]{Gangopadhyay2010}%
  \BibitemOpen
  \bibfield  {author} {\bibinfo {author} {\bibfnamefont {A.}~\bibnamefont
  {Gangopadhyay}}, \bibinfo {author} {\bibfnamefont {M.}~\bibnamefont {Dzero}},
  \ and\ \bibinfo {author} {\bibfnamefont {V.}~\bibnamefont {Galitski}},\
  }\href {\doibase 10.1103/PhysRevB.82.024303} {\bibfield  {journal} {\bibinfo
  {journal} {Phys. Rev. B}\ }\textbf {\bibinfo {volume} {82}},\ \bibinfo
  {pages} {024303} (\bibinfo {year} {2010})}\BibitemShut {NoStop}%
\bibitem [{\citenamefont {de~Clercq}\ \emph {et~al.}(2016)\citenamefont
  {de~Clercq}, \citenamefont {Oswald}, \citenamefont {Fl\"{u}hmann},
  \citenamefont {Keitch}, \citenamefont {Kienzler}, \citenamefont {Lo},
  \citenamefont {Marinelli}, \citenamefont {Nadlinger}, \citenamefont
  {Negnevitsky},\ and\ \citenamefont {Home}}]{deClercq2016}%
  \BibitemOpen
  \bibfield  {author} {\bibinfo {author} {\bibfnamefont {L.~E.}\ \bibnamefont
  {de~Clercq}}, \bibinfo {author} {\bibfnamefont {R.}~\bibnamefont {Oswald}},
  \bibinfo {author} {\bibfnamefont {C.}~\bibnamefont {Fl\"{u}hmann}}, \bibinfo
  {author} {\bibfnamefont {B.}~\bibnamefont {Keitch}}, \bibinfo {author}
  {\bibfnamefont {D.}~\bibnamefont {Kienzler}}, \bibinfo {author}
  {\bibfnamefont {H.~Y.}\ \bibnamefont {Lo}}, \bibinfo {author} {\bibfnamefont
  {M.}~\bibnamefont {Marinelli}}, \bibinfo {author} {\bibfnamefont
  {D.}~\bibnamefont {Nadlinger}}, \bibinfo {author} {\bibfnamefont
  {V.}~\bibnamefont {Negnevitsky}}, \ and\ \bibinfo {author} {\bibfnamefont
  {J.~P.}\ \bibnamefont {Home}},\ }\href {\doibase 10.1038/ncomms11218}
  {\bibfield  {journal} {\bibinfo  {journal} {Nat. Commun.}\ }\textbf {\bibinfo
  {volume} {7}} (\bibinfo {year} {2016}),\ 10.1038/ncomms11218}\BibitemShut
  {NoStop}%
\bibitem [{\citenamefont {Giscard}\ and\ \citenamefont
  {Bonhomme}(2020)}]{Giscard2020}%
  \BibitemOpen
  \bibfield  {author} {\bibinfo {author} {\bibfnamefont {P.-L.}\ \bibnamefont
  {Giscard}}\ and\ \bibinfo {author} {\bibfnamefont {C.}~\bibnamefont
  {Bonhomme}},\ }\href {\doibase 10.1103/PhysRevResearch.2.023081} {\bibfield
  {journal} {\bibinfo  {journal} {Phys. Rev. Res.}\ }\textbf {\bibinfo {volume}
  {2}},\ \bibinfo {pages} {023081} (\bibinfo {year} {2020})}\BibitemShut
  {NoStop}%
\bibitem [{\citenamefont {Bender}\ \emph {et~al.}(2007)\citenamefont {Bender},
  \citenamefont {Brody}, \citenamefont {Jones},\ and\ \citenamefont
  {Meister}}]{Bender2007}%
  \BibitemOpen
  \bibfield  {author} {\bibinfo {author} {\bibfnamefont {C.~M.}\ \bibnamefont
  {Bender}}, \bibinfo {author} {\bibfnamefont {D.~C.}\ \bibnamefont {Brody}},
  \bibinfo {author} {\bibfnamefont {H.~F.}\ \bibnamefont {Jones}}, \ and\
  \bibinfo {author} {\bibfnamefont {B.~K.}\ \bibnamefont {Meister}},\ }\href
  {\doibase 10.1103/PhysRevLett.98.040403} {\bibfield  {journal} {\bibinfo
  {journal} {Phys. Rev. Lett.}\ }\textbf {\bibinfo {volume} {98}},\ \bibinfo
  {pages} {040403} (\bibinfo {year} {2007})}\BibitemShut {NoStop}%
\bibitem [{\citenamefont {Ib\'a\~nez}\ \emph {et~al.}(2011)\citenamefont
  {Ib\'a\~nez}, \citenamefont {Mart\'{\i}nez-Garaot}, \citenamefont {Chen},
  \citenamefont {Torrontegui},\ and\ \citenamefont {Muga}}]{Ibanez2011}%
  \BibitemOpen
  \bibfield  {author} {\bibinfo {author} {\bibfnamefont {S.}~\bibnamefont
  {Ib\'a\~nez}}, \bibinfo {author} {\bibfnamefont {S.}~\bibnamefont
  {Mart\'{\i}nez-Garaot}}, \bibinfo {author} {\bibfnamefont {X.}~\bibnamefont
  {Chen}}, \bibinfo {author} {\bibfnamefont {E.}~\bibnamefont {Torrontegui}}, \
  and\ \bibinfo {author} {\bibfnamefont {J.~G.}\ \bibnamefont {Muga}},\ }\href
  {\doibase 10.1103/PhysRevA.84.023415} {\bibfield  {journal} {\bibinfo
  {journal} {Phys. Rev. A}\ }\textbf {\bibinfo {volume} {84}},\ \bibinfo
  {pages} {023415} (\bibinfo {year} {2011})}\BibitemShut {NoStop}%
\bibitem [{\citenamefont {Torosov}\ \emph {et~al.}(2013)\citenamefont
  {Torosov}, \citenamefont {Della~Valle},\ and\ \citenamefont
  {Longhi}}]{Torosov2013}%
  \BibitemOpen
  \bibfield  {author} {\bibinfo {author} {\bibfnamefont {B.~T.}\ \bibnamefont
  {Torosov}}, \bibinfo {author} {\bibfnamefont {G.}~\bibnamefont
  {Della~Valle}}, \ and\ \bibinfo {author} {\bibfnamefont {S.}~\bibnamefont
  {Longhi}},\ }\href {\doibase 10.1103/PhysRevA.87.052502} {\bibfield
  {journal} {\bibinfo  {journal} {Phys. Rev. A}\ }\textbf {\bibinfo {volume}
  {87}},\ \bibinfo {pages} {052502} (\bibinfo {year} {2013})}\BibitemShut
  {NoStop}%
\bibitem [{\citenamefont {Zheng}\ \emph {et~al.}(2013)\citenamefont {Zheng},
  \citenamefont {Hao},\ and\ \citenamefont {Long}}]{Zheng2013}%
  \BibitemOpen
  \bibfield  {author} {\bibinfo {author} {\bibfnamefont {C.}~\bibnamefont
  {Zheng}}, \bibinfo {author} {\bibfnamefont {L.}~\bibnamefont {Hao}}, \ and\
  \bibinfo {author} {\bibfnamefont {G.~L.}\ \bibnamefont {Long}},\ }\href
  {\doibase 10.1098/rsta.2012.0053} {\bibfield  {journal} {\bibinfo  {journal}
  {Phil. Trans. R. Soc. A}\ }\textbf {\bibinfo {volume} {371}},\ \bibinfo
  {pages} {20120053} (\bibinfo {year} {2013})}\BibitemShut {NoStop}%
\bibitem [{\citenamefont {Lian}\ \emph {et~al.}(2014)\citenamefont {Lian},
  \citenamefont {Zhong}, \citenamefont {Xie}, \citenamefont {Zhou},
  \citenamefont {Wu},\ and\ \citenamefont {Liao}}]{Lian2014}%
  \BibitemOpen
  \bibfield  {author} {\bibinfo {author} {\bibfnamefont {X.}~\bibnamefont
  {Lian}}, \bibinfo {author} {\bibfnamefont {H.}~\bibnamefont {Zhong}},
  \bibinfo {author} {\bibfnamefont {Q.}~\bibnamefont {Xie}}, \bibinfo {author}
  {\bibfnamefont {X.}~\bibnamefont {Zhou}}, \bibinfo {author} {\bibfnamefont
  {Y.}~\bibnamefont {Wu}}, \ and\ \bibinfo {author} {\bibfnamefont
  {W.}~\bibnamefont {Liao}},\ }\href {\doibase 10.1140/epjd/e2014-50188-1}
  {\bibfield  {journal} {\bibinfo  {journal} {Eur. Phys. J. D}\ }\textbf
  {\bibinfo {volume} {68}} (\bibinfo {year} {2014}),\
  10.1140/epjd/e2014-50188-1}\BibitemShut {NoStop}%
\bibitem [{\citenamefont {D’Ambroise}\ \emph {et~al.}(2014)\citenamefont
  {D’Ambroise}, \citenamefont {Malomed},\ and\ \citenamefont
  {Kevrekidis}}]{DAmbroise2014}%
  \BibitemOpen
  \bibfield  {author} {\bibinfo {author} {\bibfnamefont {J.}~\bibnamefont
  {D’Ambroise}}, \bibinfo {author} {\bibfnamefont {B.~A.}\ \bibnamefont
  {Malomed}}, \ and\ \bibinfo {author} {\bibfnamefont {P.~G.}\ \bibnamefont
  {Kevrekidis}},\ }\href {\doibase 10.1063/1.4883715} {\bibfield  {journal}
  {\bibinfo  {journal} {Chaos}\ }\textbf {\bibinfo {volume} {24}} (\bibinfo
  {year} {2014}),\ 10.1063/1.4883715}\BibitemShut {NoStop}%
\bibitem [{\citenamefont {Gong}\ and\ \citenamefont {Wang}(2015)}]{Gong2015}%
  \BibitemOpen
  \bibfield  {author} {\bibinfo {author} {\bibfnamefont {J.}~\bibnamefont
  {Gong}}\ and\ \bibinfo {author} {\bibfnamefont {Q.-h.}\ \bibnamefont
  {Wang}},\ }\href {\doibase 10.1103/PhysRevA.91.042135} {\bibfield  {journal}
  {\bibinfo  {journal} {Phys. Rev. A}\ }\textbf {\bibinfo {volume} {91}},\
  \bibinfo {pages} {042135} (\bibinfo {year} {2015})}\BibitemShut {NoStop}%
\bibitem [{\citenamefont {Wu}\ \emph {et~al.}(2017)\citenamefont {Wu},
  \citenamefont {Zhu}, \citenamefont {Hu}, \citenamefont {Zhou},\ and\
  \citenamefont {Zhong}}]{Wu2017}%
  \BibitemOpen
  \bibfield  {author} {\bibinfo {author} {\bibfnamefont {Y.}~\bibnamefont
  {Wu}}, \bibinfo {author} {\bibfnamefont {B.}~\bibnamefont {Zhu}}, \bibinfo
  {author} {\bibfnamefont {S.-F.}\ \bibnamefont {Hu}}, \bibinfo {author}
  {\bibfnamefont {Z.}~\bibnamefont {Zhou}}, \ and\ \bibinfo {author}
  {\bibfnamefont {H.-H.}\ \bibnamefont {Zhong}},\ }\href {\doibase
  10.1007/s11467-016-0642-x} {\bibfield  {journal} {\bibinfo  {journal} {Front.
  Phys.}\ }\textbf {\bibinfo {volume} {12}} (\bibinfo {year} {2017}),\
  10.1007/s11467-016-0642-x}\BibitemShut {NoStop}%
\bibitem [{\citenamefont {Hassan}\ \emph {et~al.}(2017)\citenamefont {Hassan},
  \citenamefont {Zhen}, \citenamefont {Solja\ifmmode \check{c}\else
  \v{c}\fi{}i\ifmmode~\acute{c}\else \'{c}\fi{}}, \citenamefont {Khajavikhan},\
  and\ \citenamefont {Christodoulides}}]{Hassan2017}%
  \BibitemOpen
  \bibfield  {author} {\bibinfo {author} {\bibfnamefont {A.~U.}\ \bibnamefont
  {Hassan}}, \bibinfo {author} {\bibfnamefont {B.}~\bibnamefont {Zhen}},
  \bibinfo {author} {\bibfnamefont {M.}~\bibnamefont {Solja\ifmmode
  \check{c}\else \v{c}\fi{}i\ifmmode~\acute{c}\else \'{c}\fi{}}}, \bibinfo
  {author} {\bibfnamefont {M.}~\bibnamefont {Khajavikhan}}, \ and\ \bibinfo
  {author} {\bibfnamefont {D.~N.}\ \bibnamefont {Christodoulides}},\ }\href
  {\doibase 10.1103/PhysRevLett.118.093002} {\bibfield  {journal} {\bibinfo
  {journal} {Phys. Rev. Lett.}\ }\textbf {\bibinfo {volume} {118}},\ \bibinfo
  {pages} {093002} (\bibinfo {year} {2017})}\BibitemShut {NoStop}%
\bibitem [{\citenamefont {Xie}\ \emph {et~al.}(2018)\citenamefont {Xie},
  \citenamefont {Rong},\ and\ \citenamefont {Liu}}]{Xie2018}%
  \BibitemOpen
  \bibfield  {author} {\bibinfo {author} {\bibfnamefont {Q.}~\bibnamefont
  {Xie}}, \bibinfo {author} {\bibfnamefont {S.}~\bibnamefont {Rong}}, \ and\
  \bibinfo {author} {\bibfnamefont {X.}~\bibnamefont {Liu}},\ }\href {\doibase
  10.1103/PhysRevA.98.052122} {\bibfield  {journal} {\bibinfo  {journal} {Phys.
  Rev. A}\ }\textbf {\bibinfo {volume} {98}},\ \bibinfo {pages} {052122}
  (\bibinfo {year} {2018})}\BibitemShut {NoStop}%
\bibitem [{\citenamefont {Pires}\ and\ \citenamefont
  {Macr\`{\i}}(2021)}]{Pires2021}%
  \BibitemOpen
  \bibfield  {author} {\bibinfo {author} {\bibfnamefont {D.~P.}\ \bibnamefont
  {Pires}}\ and\ \bibinfo {author} {\bibfnamefont {T.}~\bibnamefont
  {Macr\`{\i}}},\ }\href {\doibase 10.1103/PhysRevB.104.155141} {\bibfield
  {journal} {\bibinfo  {journal} {Phys. Rev. B}\ }\textbf {\bibinfo {volume}
  {104}},\ \bibinfo {pages} {155141} (\bibinfo {year} {2021})}\BibitemShut
  {NoStop}%
\bibitem [{\citenamefont {Dogra}\ \emph {et~al.}(2021)\citenamefont {Dogra},
  \citenamefont {Melnikov},\ and\ \citenamefont {Paraoanu}}]{Dogra2021}%
  \BibitemOpen
  \bibfield  {author} {\bibinfo {author} {\bibfnamefont {S.}~\bibnamefont
  {Dogra}}, \bibinfo {author} {\bibfnamefont {A.~A.}\ \bibnamefont {Melnikov}},
  \ and\ \bibinfo {author} {\bibfnamefont {G.~S.}\ \bibnamefont {Paraoanu}},\
  }\href {\doibase 10.1038/s42005-021-00534-2} {\bibfield  {journal} {\bibinfo
  {journal} {Commun. Phys.}\ }\textbf {\bibinfo {volume} {4}} (\bibinfo {year}
  {2021}),\ 10.1038/s42005-021-00534-2}\BibitemShut {NoStop}%
\bibitem [{\citenamefont {Xu}\ and\ \citenamefont {Guo}(2022)}]{Xu2022}%
  \BibitemOpen
  \bibfield  {author} {\bibinfo {author} {\bibfnamefont {J.}~\bibnamefont
  {Xu}}\ and\ \bibinfo {author} {\bibfnamefont {Y.}~\bibnamefont {Guo}},\
  }\href {\doibase 10.1088/1367-2630/ac69b2} {\bibfield  {journal} {\bibinfo
  {journal} {New J. Phys.}\ }\textbf {\bibinfo {volume} {24}},\ \bibinfo
  {pages} {053028} (\bibinfo {year} {2022})}\BibitemShut {NoStop}%
\bibitem [{\citenamefont {Leng}\ and\ \citenamefont {Chen}(2023)}]{Leng2023}%
  \BibitemOpen
  \bibfield  {author} {\bibinfo {author} {\bibfnamefont {Y.}~\bibnamefont
  {Leng}}\ and\ \bibinfo {author} {\bibfnamefont {F.}~\bibnamefont {Chen}},\
  }\href {\doibase 10.1088/1555-6611/ad06a1} {\bibfield  {journal} {\bibinfo
  {journal} {Laser Phys.}\ }\textbf {\bibinfo {volume} {33}},\ \bibinfo {pages}
  {125203} (\bibinfo {year} {2023})}\BibitemShut {NoStop}%
\bibitem [{\citenamefont {Kivel\"a}\ \emph {et~al.}(2024)\citenamefont
  {Kivel\"a}, \citenamefont {Dogra},\ and\ \citenamefont
  {Paraoanu}}]{Kivela2024}%
  \BibitemOpen
  \bibfield  {author} {\bibinfo {author} {\bibfnamefont {F.}~\bibnamefont
  {Kivel\"a}}, \bibinfo {author} {\bibfnamefont {S.}~\bibnamefont {Dogra}}, \
  and\ \bibinfo {author} {\bibfnamefont {G.~S.}\ \bibnamefont {Paraoanu}},\
  }\href {\doibase 10.1103/PhysRevResearch.6.023246} {\bibfield  {journal}
  {\bibinfo  {journal} {Phys. Rev. Res.}\ }\textbf {\bibinfo {volume} {6}},\
  \bibinfo {pages} {023246} (\bibinfo {year} {2024})}\BibitemShut {NoStop}%
\bibitem [{\citenamefont {Pan}\ and\ \citenamefont {Wu}(2024)}]{Pan2024}%
  \BibitemOpen
  \bibfield  {author} {\bibinfo {author} {\bibfnamefont {J.-S.}\ \bibnamefont
  {Pan}}\ and\ \bibinfo {author} {\bibfnamefont {F.}~\bibnamefont {Wu}},\
  }\href {\doibase 10.1103/PhysRevA.109.022245} {\bibfield  {journal} {\bibinfo
   {journal} {Phys. Rev. A}\ }\textbf {\bibinfo {volume} {109}},\ \bibinfo
  {pages} {022245} (\bibinfo {year} {2024})}\BibitemShut {NoStop}%
\bibitem [{\citenamefont {Jebraeilli}\ and\ \citenamefont
  {Geller}(2025)}]{Jabraeilli2025}%
  \BibitemOpen
  \bibfield  {author} {\bibinfo {author} {\bibfnamefont {A.}~\bibnamefont
  {Jebraeilli}}\ and\ \bibinfo {author} {\bibfnamefont {M.~R.}\ \bibnamefont
  {Geller}},\ }\href {\doibase 10.1103/PhysRevA.111.032211} {\bibfield
  {journal} {\bibinfo  {journal} {Phys. Rev. A}\ }\textbf {\bibinfo {volume}
  {111}},\ \bibinfo {pages} {032211} (\bibinfo {year} {2025})}\BibitemShut
  {NoStop}%
\bibitem [{\citenamefont {Liu}\ \emph {et~al.}(2025)\citenamefont {Liu},
  \citenamefont {Duan},\ and\ \citenamefont {Chen}}]{Liu2025}%
  \BibitemOpen
  \bibfield  {author} {\bibinfo {author} {\bibfnamefont {Y.}~\bibnamefont
  {Liu}}, \bibinfo {author} {\bibfnamefont {L.}~\bibnamefont {Duan}}, \ and\
  \bibinfo {author} {\bibfnamefont {Q.-H.}\ \bibnamefont {Chen}},\ }\href
  {\doibase 10.1103/PhysRevResearch.7.023004} {\bibfield  {journal} {\bibinfo
  {journal} {Phys. Rev. Res.}\ }\textbf {\bibinfo {volume} {7}},\ \bibinfo
  {pages} {023004} (\bibinfo {year} {2025})}\BibitemShut {NoStop}%
\bibitem [{\citenamefont {Fan}\ \emph {et~al.}(2025)\citenamefont {Fan},
  \citenamefont {Li}, \citenamefont {Wei}, \citenamefont {Li}, \citenamefont
  {Long}, \citenamefont {Liu}, \citenamefont {Nie}, \citenamefont {Ng},\ and\
  \citenamefont {Lu}}]{Fan2025}%
  \BibitemOpen
  \bibfield  {author} {\bibinfo {author} {\bibfnamefont {Y.-a.}\ \bibnamefont
  {Fan}}, \bibinfo {author} {\bibfnamefont {X.}~\bibnamefont {Li}}, \bibinfo
  {author} {\bibfnamefont {S.}~\bibnamefont {Wei}}, \bibinfo {author}
  {\bibfnamefont {Y.}~\bibnamefont {Li}}, \bibinfo {author} {\bibfnamefont
  {X.}~\bibnamefont {Long}}, \bibinfo {author} {\bibfnamefont {H.}~\bibnamefont
  {Liu}}, \bibinfo {author} {\bibfnamefont {X.}~\bibnamefont {Nie}}, \bibinfo
  {author} {\bibfnamefont {J.}~\bibnamefont {Ng}}, \ and\ \bibinfo {author}
  {\bibfnamefont {D.}~\bibnamefont {Lu}},\ }\href {\doibase
  10.1038/s41377-025-01769-2} {\bibfield  {journal} {\bibinfo  {journal} {Light
  Sci. Appl.}\ }\textbf {\bibinfo {volume} {14}} (\bibinfo {year} {2025}),\
  10.1038/s41377-025-01769-2}\BibitemShut {NoStop}%
\bibitem [{\citenamefont {Bender}\ and\ \citenamefont
  {Boettcher}(1998)}]{Bender1998}%
  \BibitemOpen
  \bibfield  {author} {\bibinfo {author} {\bibfnamefont {C.~M.}\ \bibnamefont
  {Bender}}\ and\ \bibinfo {author} {\bibfnamefont {S.}~\bibnamefont
  {Boettcher}},\ }\href {\doibase 10.1103/PhysRevLett.80.5243} {\bibfield
  {journal} {\bibinfo  {journal} {Phys. Rev. Lett.}\ }\textbf {\bibinfo
  {volume} {80}},\ \bibinfo {pages} {5243} (\bibinfo {year}
  {1998})}\BibitemShut {NoStop}%
\bibitem [{\citenamefont {R\"{u}ter}\ \emph {et~al.}(2010)\citenamefont
  {R\"{u}ter}, \citenamefont {Makris}, \citenamefont {El-Ganainy},
  \citenamefont {Christodoulides}, \citenamefont {Segev},\ and\ \citenamefont
  {Kip}}]{Ruter2010}%
  \BibitemOpen
  \bibfield  {author} {\bibinfo {author} {\bibfnamefont {C.~E.}\ \bibnamefont
  {R\"{u}ter}}, \bibinfo {author} {\bibfnamefont {K.~G.}\ \bibnamefont
  {Makris}}, \bibinfo {author} {\bibfnamefont {R.}~\bibnamefont {El-Ganainy}},
  \bibinfo {author} {\bibfnamefont {D.~N.}\ \bibnamefont {Christodoulides}},
  \bibinfo {author} {\bibfnamefont {M.}~\bibnamefont {Segev}}, \ and\ \bibinfo
  {author} {\bibfnamefont {D.}~\bibnamefont {Kip}},\ }\href {\doibase
  10.1038/nphys1515} {\bibfield  {journal} {\bibinfo  {journal} {Nat. Phys.}\
  }\textbf {\bibinfo {volume} {6}},\ \bibinfo {pages} {192–195} (\bibinfo
  {year} {2010})}\BibitemShut {NoStop}%
\bibitem [{\citenamefont {Liu}\ \emph {et~al.}(2024)\citenamefont {Liu},
  \citenamefont {Liu}, \citenamefont {Ni}, \citenamefont {Jia}, \citenamefont
  {Ziegler}, \citenamefont {Alù},\ and\ \citenamefont {Chen}}]{Liu2024}%
  \BibitemOpen
  \bibfield  {author} {\bibinfo {author} {\bibfnamefont {W.}~\bibnamefont
  {Liu}}, \bibinfo {author} {\bibfnamefont {Q.}~\bibnamefont {Liu}}, \bibinfo
  {author} {\bibfnamefont {X.}~\bibnamefont {Ni}}, \bibinfo {author}
  {\bibfnamefont {Y.}~\bibnamefont {Jia}}, \bibinfo {author} {\bibfnamefont
  {K.}~\bibnamefont {Ziegler}}, \bibinfo {author} {\bibfnamefont
  {A.}~\bibnamefont {Alù}}, \ and\ \bibinfo {author} {\bibfnamefont
  {F.}~\bibnamefont {Chen}},\ }\href {\doibase 10.1038/s41467-024-45226-x}
  {\bibfield  {journal} {\bibinfo  {journal} {Nat. Commun.}\ }\textbf {\bibinfo
  {volume} {15}} (\bibinfo {year} {2024}),\
  10.1038/s41467-024-45226-x}\BibitemShut {NoStop}%
\bibitem [{\citenamefont {Cooper}\ \emph {et~al.}(1995)\citenamefont {Cooper},
  \citenamefont {Khare},\ and\ \citenamefont {Sukhatme}}]{Cooper1995}%
  \BibitemOpen
  \bibfield  {author} {\bibinfo {author} {\bibfnamefont {F.}~\bibnamefont
  {Cooper}}, \bibinfo {author} {\bibfnamefont {A.}~\bibnamefont {Khare}}, \
  and\ \bibinfo {author} {\bibfnamefont {U.}~\bibnamefont {Sukhatme}},\ }\href
  {\doibase 10.1016/0370-1573(94)00080-m} {\bibfield  {journal} {\bibinfo
  {journal} {Phys. Rep.}\ }\textbf {\bibinfo {volume} {251}},\ \bibinfo {pages}
  {267–385} (\bibinfo {year} {1995})}\BibitemShut {NoStop}%
\bibitem [{\citenamefont {Fern{\'a}ndez}\ and\ \citenamefont
  {Hussin}(1999)}]{fernandez1999}%
  \BibitemOpen
  \bibfield  {author} {\bibinfo {author} {\bibfnamefont {D.~J.}\ \bibnamefont
  {Fern{\'a}ndez}}\ and\ \bibinfo {author} {\bibfnamefont {V.}~\bibnamefont
  {Hussin}},\ }\href
  {https://iopscience.iop.org/article/10.1088/0305-4470/32/19/311/meta}
  {\bibfield  {journal} {\bibinfo  {journal} {J. Phys. A: Math. Gen.}\ }\textbf
  {\bibinfo {volume} {32}},\ \bibinfo {pages} {3603} (\bibinfo {year}
  {1999})}\BibitemShut {NoStop}%
\bibitem [{\citenamefont {Mielnik}\ \emph {et~al.}(2000)\citenamefont
  {Mielnik}, \citenamefont {Nieto},\ and\ \citenamefont
  {Rosas–Ortiz}}]{Mielnik2000}%
  \BibitemOpen
  \bibfield  {author} {\bibinfo {author} {\bibfnamefont {B.}~\bibnamefont
  {Mielnik}}, \bibinfo {author} {\bibfnamefont {L.}~\bibnamefont {Nieto}}, \
  and\ \bibinfo {author} {\bibfnamefont {O.}~\bibnamefont {Rosas–Ortiz}},\
  }\href {\doibase 10.1016/s0375-9601(00)00226-7} {\bibfield  {journal}
  {\bibinfo  {journal} {Phys. Lett. A}\ }\textbf {\bibinfo {volume} {269}},\
  \bibinfo {pages} {70–78} (\bibinfo {year} {2000})}\BibitemShut {NoStop}%
\bibitem [{\citenamefont {Samsonov}\ and\ \citenamefont
  {Negro}(2004)}]{Samsonov2004}%
  \BibitemOpen
  \bibfield  {author} {\bibinfo {author} {\bibfnamefont {B.~F.}\ \bibnamefont
  {Samsonov}}\ and\ \bibinfo {author} {\bibfnamefont {J.}~\bibnamefont
  {Negro}},\ }\href {\doibase 10.1088/0305-4470/37/43/007} {\bibfield
  {journal} {\bibinfo  {journal} {J. Phy. A: Math. Ge.}\ }\textbf {\bibinfo
  {volume} {37}},\ \bibinfo {pages} {10115–10127} (\bibinfo {year}
  {2004})}\BibitemShut {NoStop}%
\bibitem [{\citenamefont {Correa}\ and\ \citenamefont
  {Plyushchay}(2007)}]{Correa2007}%
  \BibitemOpen
  \bibfield  {author} {\bibinfo {author} {\bibfnamefont {F.}~\bibnamefont
  {Correa}}\ and\ \bibinfo {author} {\bibfnamefont {M.~S.}\ \bibnamefont
  {Plyushchay}},\ }\href {\doibase 10.1016/j.aop.2006.12.002} {\bibfield
  {journal} {\bibinfo  {journal} {Ann. Phys.}\ }\textbf {\bibinfo {volume}
  {322}},\ \bibinfo {pages} {2493–2500} (\bibinfo {year} {2007})}\BibitemShut
  {NoStop}%
\bibitem [{\citenamefont {Miri}\ \emph {et~al.}(2013)\citenamefont {Miri},
  \citenamefont {Heinrich}, \citenamefont {El-Ganainy},\ and\ \citenamefont
  {Christodoulides}}]{miri2013}%
  \BibitemOpen
  \bibfield  {author} {\bibinfo {author} {\bibfnamefont {M.-A.}\ \bibnamefont
  {Miri}}, \bibinfo {author} {\bibfnamefont {M.}~\bibnamefont {Heinrich}},
  \bibinfo {author} {\bibfnamefont {R.}~\bibnamefont {El-Ganainy}}, \ and\
  \bibinfo {author} {\bibfnamefont {D.~N.}\ \bibnamefont {Christodoulides}},\
  }\href {https://journals.aps.org/prl/abstract/10.1103/PhysRevLett.110.233902}
  {\bibfield  {journal} {\bibinfo  {journal} {Phys. Rev. Lett.}\ }\textbf
  {\bibinfo {volume} {110}},\ \bibinfo {pages} {233902} (\bibinfo {year}
  {2013})}\BibitemShut {NoStop}%
\bibitem [{\citenamefont {Correa}\ \emph {et~al.}(2015)\citenamefont {Correa},
  \citenamefont {Jakubsk\'y},\ and\ \citenamefont {Plyushchay}}]{Correa2015}%
  \BibitemOpen
  \bibfield  {author} {\bibinfo {author} {\bibfnamefont {F.}~\bibnamefont
  {Correa}}, \bibinfo {author} {\bibfnamefont {V.}~\bibnamefont {Jakubsk\'y}},
  \ and\ \bibinfo {author} {\bibfnamefont {M.~S.}\ \bibnamefont {Plyushchay}},\
  }\href {\doibase 10.1103/PhysRevA.92.023839} {\bibfield  {journal} {\bibinfo
  {journal} {Phys. Rev. A}\ }\textbf {\bibinfo {volume} {92}},\ \bibinfo
  {pages} {023839} (\bibinfo {year} {2015})}\BibitemShut {NoStop}%
\bibitem [{\citenamefont {Cruz~y Cruz}\ \emph {et~al.}(2020)\citenamefont
  {Cruz~y Cruz}, \citenamefont {Razo}, \citenamefont {Rosas-Ortiz},\ and\
  \citenamefont {Zelaya}}]{CruzyCruz2020}%
  \BibitemOpen
  \bibfield  {author} {\bibinfo {author} {\bibfnamefont {S.}~\bibnamefont
  {Cruz~y Cruz}}, \bibinfo {author} {\bibfnamefont {R.}~\bibnamefont {Razo}},
  \bibinfo {author} {\bibfnamefont {O.}~\bibnamefont {Rosas-Ortiz}}, \ and\
  \bibinfo {author} {\bibfnamefont {K.}~\bibnamefont {Zelaya}},\ }\href
  {\doibase 10.1088/1402-4896/ab6525} {\bibfield  {journal} {\bibinfo
  {journal} {Phys. Scr.}\ }\textbf {\bibinfo {volume} {95}},\ \bibinfo {pages}
  {044009} (\bibinfo {year} {2020})}\BibitemShut {NoStop}%
\bibitem [{\citenamefont {Garc{\'i}a-Mu{\~n}oz}\ \emph
  {et~al.}(2023)\citenamefont {Garc{\'i}a-Mu{\~n}oz}, \citenamefont {Raya},\
  and\ \citenamefont {Concha-S{\'a}nchez}}]{Garcia2023}%
  \BibitemOpen
  \bibfield  {author} {\bibinfo {author} {\bibfnamefont {J.~D.}\ \bibnamefont
  {Garc{\'i}a-Mu{\~n}oz}}, \bibinfo {author} {\bibfnamefont {A.}~\bibnamefont
  {Raya}}, \ and\ \bibinfo {author} {\bibfnamefont {Y.}~\bibnamefont
  {Concha-S{\'a}nchez}},\ }\href
  {https://iopscience.iop.org/article/10.1088/1402-4896/ad05ac/meta} {\bibfield
   {journal} {\bibinfo  {journal} {Phys. Scr.}\ }\textbf {\bibinfo {volume}
  {98}},\ \bibinfo {pages} {125203} (\bibinfo {year} {2023})}\BibitemShut
  {NoStop}%
\bibitem [{\citenamefont {Zu{\~n}iga-Segundo}\ \emph
  {et~al.}(2014)\citenamefont {Zu{\~n}iga-Segundo}, \citenamefont
  {Rodr{\'\i}guez-Lara}, \citenamefont {Fern{\`a}ndez},\ and\ \citenamefont
  {Moya-Cessa}}]{Zuniga2014}%
  \BibitemOpen
  \bibfield  {author} {\bibinfo {author} {\bibfnamefont {A.}~\bibnamefont
  {Zu{\~n}iga-Segundo}}, \bibinfo {author} {\bibfnamefont {B.}~\bibnamefont
  {Rodr{\'\i}guez-Lara}}, \bibinfo {author} {\bibfnamefont {D.~J.}\
  \bibnamefont {Fern{\`a}ndez}}, \ and\ \bibinfo {author} {\bibfnamefont
  {H.}~\bibnamefont {Moya-Cessa}},\ }\href
  {https://opg.optica.org/oe/abstract.cfm?URI=oe-22-1-987} {\bibfield
  {journal} {\bibinfo  {journal} {Opt. Express.}\ }\textbf {\bibinfo {volume}
  {22}},\ \bibinfo {pages} {987} (\bibinfo {year} {2014})}\BibitemShut
  {NoStop}%
\bibitem [{\citenamefont {Schulze-Halberg}\ and\ \citenamefont
  {Roy}(2017)}]{Schulze2017}%
  \BibitemOpen
  \bibfield  {author} {\bibinfo {author} {\bibfnamefont {A.}~\bibnamefont
  {Schulze-Halberg}}\ and\ \bibinfo {author} {\bibfnamefont {P.}~\bibnamefont
  {Roy}},\ }\href
  {https://iopscience.iop.org/article/10.1088/1751-8121/aa8249/meta} {\bibfield
   {journal} {\bibinfo  {journal} {J. Phys. A. Math. Theor.}\ }\textbf
  {\bibinfo {volume} {50}},\ \bibinfo {pages} {365205} (\bibinfo {year}
  {2017})}\BibitemShut {NoStop}%
\bibitem [{\citenamefont {Contreras-Astorga}\ and\ \citenamefont
  {Jakubsk{\`y}}(2019)}]{Contreras2019}%
  \BibitemOpen
  \bibfield  {author} {\bibinfo {author} {\bibfnamefont {A.}~\bibnamefont
  {Contreras-Astorga}}\ and\ \bibinfo {author} {\bibfnamefont {V.}~\bibnamefont
  {Jakubsk{\`y}}},\ }\href
  {https://journals.aps.org/pra/abstract/10.1103/PhysRevA.99.053812} {\bibfield
   {journal} {\bibinfo  {journal} {Phys. Rev. A}\ }\textbf {\bibinfo {volume}
  {99}},\ \bibinfo {pages} {053812} (\bibinfo {year} {2019})}\BibitemShut
  {NoStop}%
\bibitem [{\citenamefont {Maldonado-Villamizar}\ \emph
  {et~al.}(2021)\citenamefont {Maldonado-Villamizar}, \citenamefont
  {Gonz\'alez-Guti\'errez}, \citenamefont {Villanueva-Vergara},\ and\
  \citenamefont {Rodr\'iguez-Lara}}]{Maldonado2021}%
  \BibitemOpen
  \bibfield  {author} {\bibinfo {author} {\bibfnamefont {F.~H.}\ \bibnamefont
  {Maldonado-Villamizar}}, \bibinfo {author} {\bibfnamefont {C.~A.}\
  \bibnamefont {Gonz\'alez-Guti\'errez}}, \bibinfo {author} {\bibfnamefont
  {L.}~\bibnamefont {Villanueva-Vergara}}, \ and\ \bibinfo {author}
  {\bibfnamefont {B.~M.}\ \bibnamefont {Rodr\'iguez-Lara}},\ }\href {\doibase
  10.1038/s41598-021-95259-1} {\bibfield  {journal} {\bibinfo  {journal} {Sci.
  Rep.}\ }\textbf {\bibinfo {volume} {11}},\ \bibinfo {pages} {16467} (\bibinfo
  {year} {2021})}\BibitemShut {NoStop}%
\bibitem [{\citenamefont {Kafuri}\ \emph {et~al.}(2024)\citenamefont {Kafuri},
  \citenamefont {Maldonado-Villamizar}, \citenamefont {Moroz},\ and\
  \citenamefont {Rodríguez-Lara}}]{Kafuri2024}%
  \BibitemOpen
  \bibfield  {author} {\bibinfo {author} {\bibfnamefont {A.}~\bibnamefont
  {Kafuri}}, \bibinfo {author} {\bibfnamefont {F.~H.}\ \bibnamefont
  {Maldonado-Villamizar}}, \bibinfo {author} {\bibfnamefont {A.}~\bibnamefont
  {Moroz}}, \ and\ \bibinfo {author} {\bibfnamefont {B.~M.}\ \bibnamefont
  {Rodríguez-Lara}},\ }\href {\doibase 10.1364/josab.522504} {\bibfield
  {journal} {\bibinfo  {journal} {J. Opt. Soc. Am. B}\ }\textbf {\bibinfo
  {volume} {41}},\ \bibinfo {pages} {C82} (\bibinfo {year} {2024})}\BibitemShut
  {NoStop}%
\bibitem [{\citenamefont {Bocanegra-Garay}\ \emph
  {et~al.}(2024{\natexlab{a}})\citenamefont {Bocanegra-Garay}, \citenamefont
  {Hernández-Sánchez}, \citenamefont {Ramos-Prieto}, \citenamefont
  {Soto-Eguibar},\ and\ \citenamefont {Moya-Cessa}}]{BocanegraGaray2024}%
  \BibitemOpen
  \bibfield  {author} {\bibinfo {author} {\bibfnamefont {I.~A.}\ \bibnamefont
  {Bocanegra-Garay}}, \bibinfo {author} {\bibfnamefont {L.}~\bibnamefont
  {Hernández-Sánchez}}, \bibinfo {author} {\bibfnamefont {I.}~\bibnamefont
  {Ramos-Prieto}}, \bibinfo {author} {\bibfnamefont {F.}~\bibnamefont
  {Soto-Eguibar}}, \ and\ \bibinfo {author} {\bibfnamefont {H.~M.}\
  \bibnamefont {Moya-Cessa}},\ }\href {\doibase 10.1088/1402-4896/ad20bd}
  {\bibfield  {journal} {\bibinfo  {journal} {Phys. Scr.}\ }\textbf {\bibinfo
  {volume} {99}},\ \bibinfo {pages} {035216} (\bibinfo {year}
  {2024}{\natexlab{a}})}\BibitemShut {NoStop}%
\bibitem [{\citenamefont {Bocanegra-Garay}\ \emph
  {et~al.}(2024{\natexlab{b}})\citenamefont {Bocanegra-Garay}, \citenamefont
  {Castillo-Celeita}, \citenamefont {Negro}, \citenamefont {Nieto},\ and\
  \citenamefont {G\'omez-Ruiz}}]{Bocanegra_2024PRR}%
  \BibitemOpen
  \bibfield  {author} {\bibinfo {author} {\bibfnamefont {I.~A.}\ \bibnamefont
  {Bocanegra-Garay}}, \bibinfo {author} {\bibfnamefont {M.}~\bibnamefont
  {Castillo-Celeita}}, \bibinfo {author} {\bibfnamefont {J.}~\bibnamefont
  {Negro}}, \bibinfo {author} {\bibfnamefont {L.~M.}\ \bibnamefont {Nieto}}, \
  and\ \bibinfo {author} {\bibfnamefont {F.~J.}\ \bibnamefont {G\'omez-Ruiz}},\
  }\href {\doibase 10.1103/PhysRevResearch.6.043218} {\bibfield  {journal}
  {\bibinfo  {journal} {Phys. Rev. Res.}\ }\textbf {\bibinfo {volume} {6}},\
  \bibinfo {pages} {043218} (\bibinfo {year} {2024}{\natexlab{b}})}\BibitemShut
  {NoStop}%
\bibitem [{\citenamefont {Rosas-Ortiz}\ \emph {et~al.}(2015)\citenamefont
  {Rosas-Ortiz}, \citenamefont {Castaños},\ and\ \citenamefont
  {Schuch}}]{RosasOrtiz2015}%
  \BibitemOpen
  \bibfield  {author} {\bibinfo {author} {\bibfnamefont {O.}~\bibnamefont
  {Rosas-Ortiz}}, \bibinfo {author} {\bibfnamefont {O.}~\bibnamefont
  {Castaños}}, \ and\ \bibinfo {author} {\bibfnamefont {D.}~\bibnamefont
  {Schuch}},\ }\href {\doibase 10.1088/1751-8113/48/44/445302} {\bibfield
  {journal} {\bibinfo  {journal} {J. Phys. A. Math. Theor.}\ }\textbf {\bibinfo
  {volume} {48}},\ \bibinfo {pages} {445302} (\bibinfo {year}
  {2015})}\BibitemShut {NoStop}%
\bibitem [{\citenamefont {Rosas-Ortiz}\ and\ \citenamefont
  {Zelaya}(2018)}]{RosasOrtiz2018}%
  \BibitemOpen
  \bibfield  {author} {\bibinfo {author} {\bibfnamefont {O.}~\bibnamefont
  {Rosas-Ortiz}}\ and\ \bibinfo {author} {\bibfnamefont {K.}~\bibnamefont
  {Zelaya}},\ }\href {\doibase 10.1016/j.aop.2017.10.020} {\bibfield  {journal}
  {\bibinfo  {journal} {Ann. Phys.}\ }\textbf {\bibinfo {volume} {388}},\
  \bibinfo {pages} {26–53} (\bibinfo {year} {2018})}\BibitemShut {NoStop}%
\bibitem [{\citenamefont {Zelaya}\ \emph {et~al.}(2020)\citenamefont {Zelaya},
  \citenamefont {Cruz~y Cruz},\ and\ \citenamefont {Rosas-Ortiz}}]{Zelaya2020}%
  \BibitemOpen
  \bibfield  {author} {\bibinfo {author} {\bibfnamefont {K.}~\bibnamefont
  {Zelaya}}, \bibinfo {author} {\bibfnamefont {S.}~\bibnamefont {Cruz~y Cruz}},
  \ and\ \bibinfo {author} {\bibfnamefont {O.}~\bibnamefont {Rosas-Ortiz}},\
  }\enquote {\bibinfo {title} {On the construction of non-hermitian
  hamiltonians with all-real spectra through supersymmetric algorithms},}\ in\
  \href {\doibase 10.1007/978-3-030-53305-2_18} {\emph {\bibinfo {booktitle}
  {Geometric Methods in Physics XXXVIII}}}\ (\bibinfo  {publisher} {Springer
  International Publishing},\ \bibinfo {year} {2020})\ p.\ \bibinfo {pages}
  {283–292}\BibitemShut {NoStop}%
\bibitem [{\citenamefont {Bocanegra}\ and\ \citenamefont {Cruz~y
  Cruz}(2022)}]{Bocanegra2022}%
  \BibitemOpen
  \bibfield  {author} {\bibinfo {author} {\bibfnamefont {I.}~\bibnamefont
  {Bocanegra}}\ and\ \bibinfo {author} {\bibfnamefont {S.}~\bibnamefont {Cruz~y
  Cruz}},\ }\href {\doibase 10.3390/sym14030432} {\bibfield  {journal}
  {\bibinfo  {journal} {Symmetry}\ }\textbf {\bibinfo {volume} {14}},\ \bibinfo
  {pages} {432} (\bibinfo {year} {2022})}\BibitemShut {NoStop}%
\bibitem [{\citenamefont {Bender}\ and\ \citenamefont
  {Milton}(1998)}]{Bender1998SUSY}%
  \BibitemOpen
  \bibfield  {author} {\bibinfo {author} {\bibfnamefont {C.~M.}\ \bibnamefont
  {Bender}}\ and\ \bibinfo {author} {\bibfnamefont {K.~A.}\ \bibnamefont
  {Milton}},\ }\href {\doibase 10.1103/PhysRevD.57.3595} {\bibfield  {journal}
  {\bibinfo  {journal} {Phys. Rev. D}\ }\textbf {\bibinfo {volume} {57}},\
  \bibinfo {pages} {3595} (\bibinfo {year} {1998})}\BibitemShut {NoStop}%
\bibitem [{\citenamefont {Bloch}(1946)}]{Bloch1946}%
  \BibitemOpen
  \bibfield  {author} {\bibinfo {author} {\bibfnamefont {F.}~\bibnamefont
  {Bloch}},\ }\href {\doibase 10.1103/PhysRev.70.460} {\bibfield  {journal}
  {\bibinfo  {journal} {Phys. Rev.}\ }\textbf {\bibinfo {volume} {70}},\
  \bibinfo {pages} {460} (\bibinfo {year} {1946})}\BibitemShut {NoStop}%
\bibitem [{\citenamefont {Silver}\ \emph
  {et~al.}(1984{\natexlab{a}})\citenamefont {Silver}, \citenamefont {Joseph},
  \citenamefont {Chen}, \citenamefont {Sank},\ and\ \citenamefont
  {Hoult}}]{Silver1984PRA}%
  \BibitemOpen
  \bibfield  {author} {\bibinfo {author} {\bibfnamefont {M.~S.}\ \bibnamefont
  {Silver}}, \bibinfo {author} {\bibfnamefont {R.~I.}\ \bibnamefont {Joseph}},
  \bibinfo {author} {\bibfnamefont {C.-N.}\ \bibnamefont {Chen}}, \bibinfo
  {author} {\bibfnamefont {V.~J.}\ \bibnamefont {Sank}}, \ and\ \bibinfo
  {author} {\bibfnamefont {D.~I.}\ \bibnamefont {Hoult}},\ }\href {\doibase
  10.1038/310681a0} {\bibfield  {journal} {\bibinfo  {journal} {Nature}\
  }\textbf {\bibinfo {volume} {310}},\ \bibinfo {pages} {681–683} (\bibinfo
  {year} {1984}{\natexlab{a}})}\BibitemShut {NoStop}%
\bibitem [{\citenamefont {Silver}\ \emph
  {et~al.}(1984{\natexlab{b}})\citenamefont {Silver}, \citenamefont {Joseph},\
  and\ \citenamefont {Hoult}}]{Silver1984JMR}%
  \BibitemOpen
  \bibfield  {author} {\bibinfo {author} {\bibfnamefont {M.}~\bibnamefont
  {Silver}}, \bibinfo {author} {\bibfnamefont {R.}~\bibnamefont {Joseph}}, \
  and\ \bibinfo {author} {\bibfnamefont {D.}~\bibnamefont {Hoult}},\ }\href
  {\doibase 10.1016/0022-2364(84)90181-1} {\bibfield  {journal} {\bibinfo
  {journal} {J. Magn. Reson.}\ }\textbf {\bibinfo {volume} {59}},\ \bibinfo
  {pages} {347–351} (\bibinfo {year} {1984}{\natexlab{b}})}\BibitemShut
  {NoStop}%
\bibitem [{\citenamefont {Silver}\ \emph {et~al.}(1985)\citenamefont {Silver},
  \citenamefont {Joseph},\ and\ \citenamefont {Hoult}}]{Silver1985}%
  \BibitemOpen
  \bibfield  {author} {\bibinfo {author} {\bibfnamefont {M.~S.}\ \bibnamefont
  {Silver}}, \bibinfo {author} {\bibfnamefont {R.~I.}\ \bibnamefont {Joseph}},
  \ and\ \bibinfo {author} {\bibfnamefont {D.~I.}\ \bibnamefont {Hoult}},\
  }\href {\doibase 10.1103/PhysRevA.31.2753} {\bibfield  {journal} {\bibinfo
  {journal} {Phys. Rev. A}\ }\textbf {\bibinfo {volume} {31}},\ \bibinfo
  {pages} {2753} (\bibinfo {year} {1985})}\BibitemShut {NoStop}%
\bibitem [{\citenamefont {Mart\'{\i}nez-Tibaduiza}\ \emph
  {et~al.}(2023)\citenamefont {Mart\'{\i}nez-Tibaduiza}, \citenamefont
  {Gonz\'alez-Arciniegas}, \citenamefont {Farina}, \citenamefont
  {Cavalcanti-Duriez},\ and\ \citenamefont {Khoury}}]{Martinez2023}%
  \BibitemOpen
  \bibfield  {author} {\bibinfo {author} {\bibfnamefont {D.}~\bibnamefont
  {Mart\'{\i}nez-Tibaduiza}}, \bibinfo {author} {\bibfnamefont
  {C.}~\bibnamefont {Gonz\'alez-Arciniegas}}, \bibinfo {author} {\bibfnamefont
  {C.}~\bibnamefont {Farina}}, \bibinfo {author} {\bibfnamefont
  {A.}~\bibnamefont {Cavalcanti-Duriez}}, \ and\ \bibinfo {author}
  {\bibfnamefont {A.~Z.}\ \bibnamefont {Khoury}},\ }\href {\doibase
  10.1103/PhysRevA.107.042211} {\bibfield  {journal} {\bibinfo  {journal}
  {Phys. Rev. A}\ }\textbf {\bibinfo {volume} {107}},\ \bibinfo {pages}
  {042211} (\bibinfo {year} {2023})}\BibitemShut {NoStop}%
\bibitem [{\citenamefont {Enríquez}\ and\ \citenamefont {Cruz~y
  Cruz}(2018)}]{Enriquez2018}%
  \BibitemOpen
  \bibfield  {author} {\bibinfo {author} {\bibfnamefont {M.}~\bibnamefont
  {Enríquez}}\ and\ \bibinfo {author} {\bibfnamefont {S.}~\bibnamefont {Cruz~y
  Cruz}},\ }\href {\doibase 10.3390/sym10110567} {\bibfield  {journal}
  {\bibinfo  {journal} {Symmetry}\ }\textbf {\bibinfo {volume} {10}},\ \bibinfo
  {pages} {567} (\bibinfo {year} {2018})}\BibitemShut {NoStop}%
\bibitem [{\citenamefont {Liu}\ \emph {et~al.}(2020)\citenamefont {Liu},
  \citenamefont {Li}, \citenamefont {Wang}, \citenamefont {Ke}, \citenamefont
  {Qin}, \citenamefont {Wang}, \citenamefont {Liu}, \citenamefont {Gao},
  \citenamefont {Berini},\ and\ \citenamefont {Lu}}]{Liu2020}%
  \BibitemOpen
  \bibfield  {author} {\bibinfo {author} {\bibfnamefont {Q.}~\bibnamefont
  {Liu}}, \bibinfo {author} {\bibfnamefont {S.}~\bibnamefont {Li}}, \bibinfo
  {author} {\bibfnamefont {B.}~\bibnamefont {Wang}}, \bibinfo {author}
  {\bibfnamefont {S.}~\bibnamefont {Ke}}, \bibinfo {author} {\bibfnamefont
  {C.}~\bibnamefont {Qin}}, \bibinfo {author} {\bibfnamefont {K.}~\bibnamefont
  {Wang}}, \bibinfo {author} {\bibfnamefont {W.}~\bibnamefont {Liu}}, \bibinfo
  {author} {\bibfnamefont {D.}~\bibnamefont {Gao}}, \bibinfo {author}
  {\bibfnamefont {P.}~\bibnamefont {Berini}}, \ and\ \bibinfo {author}
  {\bibfnamefont {P.}~\bibnamefont {Lu}},\ }\href {\doibase
  10.1103/PhysRevLett.124.153903} {\bibfield  {journal} {\bibinfo  {journal}
  {Phys. Rev. Lett.}\ }\textbf {\bibinfo {volume} {124}},\ \bibinfo {pages}
  {153903} (\bibinfo {year} {2020})}\BibitemShut {NoStop}%
\bibitem [{\citenamefont {Luo}\ \emph {et~al.}(2013)\citenamefont {Luo},
  \citenamefont {Huang}, \citenamefont {Zhong}, \citenamefont {Qin},
  \citenamefont {Xie}, \citenamefont {Kivshar},\ and\ \citenamefont
  {Lee}}]{Luo2013}%
  \BibitemOpen
  \bibfield  {author} {\bibinfo {author} {\bibfnamefont {X.}~\bibnamefont
  {Luo}}, \bibinfo {author} {\bibfnamefont {J.}~\bibnamefont {Huang}}, \bibinfo
  {author} {\bibfnamefont {H.}~\bibnamefont {Zhong}}, \bibinfo {author}
  {\bibfnamefont {X.}~\bibnamefont {Qin}}, \bibinfo {author} {\bibfnamefont
  {Q.}~\bibnamefont {Xie}}, \bibinfo {author} {\bibfnamefont {Y.~S.}\
  \bibnamefont {Kivshar}}, \ and\ \bibinfo {author} {\bibfnamefont
  {C.}~\bibnamefont {Lee}},\ }\href {\doibase 10.1103/PhysRevLett.110.243902}
  {\bibfield  {journal} {\bibinfo  {journal} {Phys. Rev. Lett.}\ }\textbf
  {\bibinfo {volume} {110}},\ \bibinfo {pages} {243902} (\bibinfo {year}
  {2013})}\BibitemShut {NoStop}%
\bibitem [{\citenamefont {Cornelius}\ \emph {et~al.}(2022)\citenamefont
  {Cornelius}, \citenamefont {Xu}, \citenamefont {Saxena}, \citenamefont
  {Chenu},\ and\ \citenamefont {del Campo}}]{Cornelius2022}%
  \BibitemOpen
  \bibfield  {author} {\bibinfo {author} {\bibfnamefont {J.}~\bibnamefont
  {Cornelius}}, \bibinfo {author} {\bibfnamefont {Z.}~\bibnamefont {Xu}},
  \bibinfo {author} {\bibfnamefont {A.}~\bibnamefont {Saxena}}, \bibinfo
  {author} {\bibfnamefont {A.}~\bibnamefont {Chenu}}, \ and\ \bibinfo {author}
  {\bibfnamefont {A.}~\bibnamefont {del Campo}},\ }\href {\doibase
  10.1103/PhysRevLett.128.190402} {\bibfield  {journal} {\bibinfo  {journal}
  {Phys. Rev. Lett.}\ }\textbf {\bibinfo {volume} {128}},\ \bibinfo {pages}
  {190402} (\bibinfo {year} {2022})}\BibitemShut {NoStop}%
\end{thebibliography}%

\end{document}